\documentclass[aps,pra,showpacs,superscriptaddress,amssymb,twocolumn]{revtex4-1}
\usepackage{graphicx,appendix,hyperref,bm,amsmath,amsfonts,dcolumn}
\usepackage[caption=false]{subfig}

\DeclareGraphicsExtensions{.pdf,.eps,.png,.jpg,.mps}
\newcommand{\bra}[1]{\left\langle #1 \right|}
\newcommand{\ket}[1]{\left|#1\right\rangle}

\def\BEq{\begin{equation}}
\def\EEq{\end{equation}}
\def\BEqA{\begin{eqnarray}}
\def\EEqA{\end{eqnarray}}
\def\BW{\begin{widetext}}
\def\EW{\end{widetext}}

\begin{document}


\title{High-fidelity CZ gate for resonator-based superconducting quantum computers}

\author{Joydip Ghosh}
\email{joydip.ghosh@gmail.com}
\affiliation{Department of Physics and Astronomy, University of Georgia, Athens, Georgia 30602, USA}

\author{Andrei Galiautdinov}
\affiliation{Department of Physics and Astronomy, University of Georgia, Athens, Georgia 30602, USA}
\affiliation{Department of Electrical Engineering, University of California, Riverside, California 92521, USA}

\author{Zhongyuan Zhou}
\affiliation{Department of Physics and Astronomy, University of Georgia, Athens, Georgia 30602, USA}

\author{Alexander N. Korotkov}
\affiliation{Department of Electrical Engineering, University of California, Riverside, California 92521, USA}

\author{John M. Martinis}
\affiliation{Department of Physics, University of California, Santa Barbara, California 93106, USA}

\author{Michael R. Geller}
\email{mgeller@uga.edu}
\affiliation{Department of Physics and Astronomy, University of Georgia, Athens, Georgia 30602, USA}

\date{\today}

\begin{abstract}
A possible building block for a scalable quantum computer has recently been demonstrated [M. Mariantoni {\it et al.}, Science {\bf 334}, 61 (2011)]. This architecture consists of superconducting qubits capacitively coupled both to individual memory resonators as well as a common bus. In this work we study a natural primitive entangling gate for this and related resonator-based architectures, which consists of a controlled-$\sigma^z$ ({\sf CZ}) operation between a qubit and the bus. The {\sf CZ} gate is implemented with the aid of the non-computational qubit $|2\rangle$ state [F. W. Strauch, {\it et al.}, Phys. Rev. Lett. {\bf 91}, 167005 (2003)]. Assuming phase or transmon qubits with $300 \, {\rm MHz}$ anharmonicity, we show that by using only low frequency qubit-bias control it is possible to implement the qubit-bus {\sf CZ} gate with 99.9\% (99.99\%) fidelity in about $17 \, {\rm ns}$ ($23 \, {\rm ns}$) with a realistic two-parameter pulse profile, plus two auxiliary $z$ rotations. The fidelity measure we refer to here is a state-averaged intrinsic process fidelity, which does not include any effects of noise or decoherence. These results apply to a multi-qubit device that includes strongly coupled memory resonators. We investigate the performance of the qubit-bus {\sf CZ} gate as a function of qubit anharmonicity, indentify the dominant intrinsic error mechanism and derive an associated fidelity estimator, quantify the pulse shape sensitivity and precision requirements, simulate qubit-qubit {\sf CZ} gates that are mediated by the bus resonator, and also attempt a global optimization of system parameters including resonator frequencies and couplings. Our results are relevant for a wide range of superconducting hardware designs that incorporate resonators and suggest that it should be possible to demonstrate a $99.9\%$ {\sf CZ} gate with existing transmon qubits, which would constitute an important step towards the development of an error-corrected superconducting quantum computer.
\end{abstract}


\maketitle

\section{QVN ARCHITECTURE}
\label{sec:INTRO}

Reaching the fidelity threshold for fault-tolerant quantum computation with superconducting electrical circuits \cite{YouPT05,SchoelkopfNat08,ClarkeNat08} will probably require improvement in three areas: qubit coherence, readout, and qubit-qubit coupling tunability. Fortunately, the coherence times of superconducting transmon qubits \cite{KochPRA07,HouckQIP09} have increased dramatically, exceeding $10 \mu s$ in the three-dimensional version \cite{PaikPRL11,RigettiPRB12}. Fast, threshold-fidelity nondestructive measurement has not yet been demonstrated, but is being actively pursued \cite{MalletNatPhys09,JohnsonNatPhys10,BoissonneaultPRL10,ReedPRL10,JohnsonPRL12}. Some method for turning off the interaction between device elements---beyond simple frequency detuning---is also desirable for high-fidelity operations. A variety of tunable coupling circuits have been demonstrated \cite{HimeSci06,vanderPloegPRL07,NiskanenSci07,HarrisPRL07,AllmanPRL10,BialczakPRL11,GroszkowskiPRB11}, but these considerably increase the complexity of the hardware and it is not clear whether they will be practical for large scale implementation. The coupling can also be controlled by the application of microwave pulses  \cite{YouPRB05,RigettiPRL05,LiuPRL06,ParaoanuPRB06,AshhabPRB06,GrajcarPRB06,AshhabPRB07,LiuPRB07,RigettiPRB10,deGrootNatPhys10,ChowPRL11,deGrootNJP12,ChowPRL12}. 

\begin{figure}[htb]
\centering
\includegraphics[angle=0,width=1.00\linewidth]{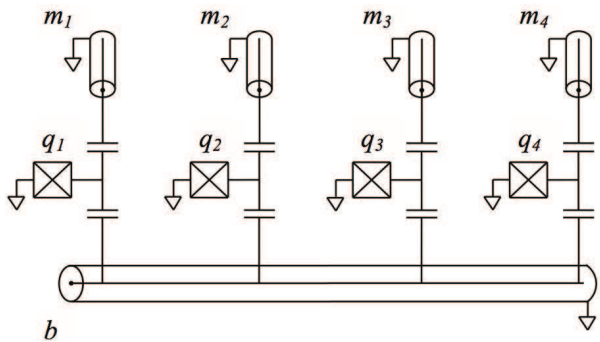}
\caption{Layout of the four-qubit QVN processor. The $q_{i}$ represent
superconducting qubits capacitively coupled to memory resonators $m_{i}$ 
as well as a resonator bus $b$.}
\label{fig:rezqu}
\end{figure}

An alternative approach has been introduced by Mariantoni {\it et al.}~\cite{MariantoniSci11} and theoretically analyzed in Ref.~\cite{GaliautdinovPRA12}. In this {\it quantum von Neumann} (QVN) architecture, superconducting qubits are capacitively coupled both to individual memory resonators as well as a common bus, as illustrated in Fig.~\ref{fig:rezqu}. The crossed boxes in Fig.~\ref{fig:rezqu} represent the phase qubits \cite{MartinisQIP09} employed by Mariantoni {\it et al.}~\cite{MariantoniSci11}, however other qubit designs such as the transmon may be used here as well. The key feature of this architecture is that information (data) is stored in memory resonators that are isolated by {\it two} detuned coupling steps from the bus. Qubits are used to transfer information to and from the bus or entangle with it, and to implement single-qubit operations, but are otherwise kept in their ground states. No more than one qubit (attached to the same bus) is to be occupied at any time. Such an approach significantly improves the effective on/off ratio without introducing the added complexity of nonlinear tunable coupling circuitry. The spectral crowding problem of the usual qubit-bus architecture is greatly reduced because the four-step coupling between memory resonators is negligible. And an added benefit of the QVN approach is that the longer coherence times of the memory elements reduce the overall decoherence rate of the device. (In Ref.~\cite{GaliautdinovPRA12}, the architecture we consider is referred to as the resonator-zero-qubit architecture, but here we will follow the QVN terminology of \cite{MariantoniSci11}.)

The QVN architecture of Mariantoni {\it et al.}~\cite{MariantoniSci11} is not, by itself, capable of large-scale, fault-tolerant quantum computation, nor is it known how multiple QVN devices might be integrated into a scalable design. The problem of designing scalable, fault-tolerant architectures for superconducting qubits is of great interest and importance \cite{DiVincenzoPhysScr09,GhoshFowler&GellerPRA12}, but is still in its infancy. We expect the gate design approach discussed here to be applicable to future architecture designs incorporating qubits coupled to resonators, and perhaps beyond.

Along with high-fidelity single-qubit rotations \cite{MotzoiPRL09,GambettaPRA11}, quantum computing with the QVN processor requires two additional types of operations. The first is state transfer between the different physical components, which has to be performed frequently during a computation. The simplest case of state transfer is between a qubit and its associated memory (or the reverse). This case is investigated in Ref.~\cite{GaliautdinovPRA12}, where two important observations are made: First, in contrast with the usual {\sf SWAP} or i{\sf SWAP} operation, which must be able to simultaneously transfer quantum information in two directions, only unidirectional state transfers are required in the QVN system. This is because adjacent qubits and resonators are---by agreement---never simultaneously populated. Second, the phase of a transferred $|1\rangle$ state is immaterial, as it can always be adjusted by future qubit $z$ rotations \cite{HofheinzNat08,HofheinzNat09}. These two simplifications allow the resulting state transfer operation, called a {\sf MOVE} gate, to be carried out with extremely high intrinsic fidelity---perfectly for a truncated model---with a simple four-parameter pulse profile. By intrinsic fidelity we mean the process fidelity (defined below) in the absence of noise or decoherence. The need for four control parameters immediately follows from the requirement that after a {\sf MOVE} gate, the probability amplitudes must vanish on two device components, the component ($q$ or $m$) the state is leaving, and the bus $b$. Each zero imposes two real parameters, and no other probability amplitudes acquire weight (in the truncated model). Fixing the phase of  the {\sf MOVE} gate, if desired, requires one additional control parameter in the form of a local $z$ rotational angle.

\begin{figure}[htb]
\centering
\includegraphics[angle=0,width=1.1\linewidth]{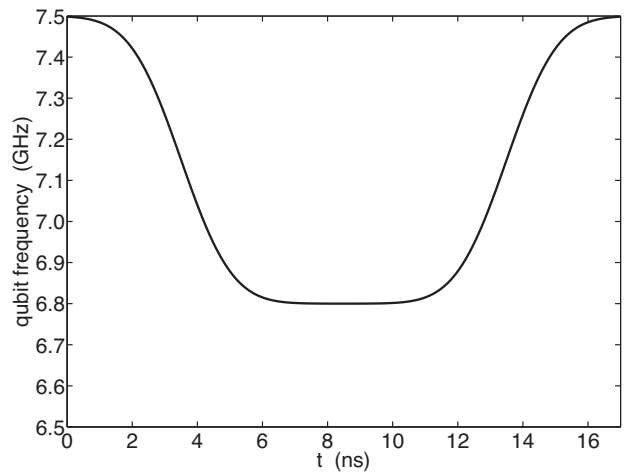}
\caption{Two-parameter CZ pulse profile (\ref{pulseEq}) for the case of $\omega_{\rm on}/2 \pi=6.8 \, {\rm GHz}$, $\omega_{\rm off}/2 \pi=7.5 \, {\rm GHz}$, $t_{\rm ramp}=7 \, {\rm ns}$, $\sigma= 1.24 \, {\rm ns}$, and $t_{\rm on}=10 \, {\rm ns}$. The total gate time excluding auxiliary $z$ rotations is  $t_{\rm gate}= 17 \, {\rm ns}$. The example shown is representative of a 99.9\% fidelity gate for a qubit with $300 \, {\rm MHz}$ anharmonicity.}
\label{fig:pulse}
\end{figure}

State transfer between a qubit and the bus (or the reverse) can be analyzed in the same way, although in this case more pulse-shape parameters are required. In an $n$-qubit QVN processor (consisting of $n$ qubits, $n$ memory resonators, and the bus), zero-amplitude conditions must be enforced on the additional $n-1$ qubits, leading to a total of $2(n+1)$  pulse parameters, plus one $z$ rotation angle. This makes quasi-exact state transfer to and from the bus a considerably more challenging operation. Simpler three-parameter approximate transfers, however, can still be implemented with very high fidelity, even when the coupling is strong (see below).

Quantum computing with the QVN system also requires a universal SU(4) entangling gate, the most natural being a controlled-$\sigma^z$ ({\sf CZ}) operation between a qubit and the bus. The {\sf CZ} gate investigated here makes essential use of the non-computational qubit $|2\rangle$ state and was first proposed by Strauch and coworkers \cite{StrauchPRL03}. The Strauch gate has been investigated by many authors and has been demonstrated in several systems \cite{DiCarloNat09,HaackPRB10,DiCarloNat10,YamamotoPRB10,GaliautdinovArxiv1103.4641,LuPre12}. 

The present paper extends previous work by considering device parameters and pulse shapes appropriate for the QVN system, and by optimizing the {\sf CZ} fidelity in a multi-qubit device. We are especially interested in whether the absence of an active tunable coupler results in any significant limitation on the obtainable fidelity, given a reasonable amount of qubit coherence, and whether very high intrinsic fidelities can be reached with a simple and experimentally realistic (filtered rectangular) pulse shape. We also study how the gate performance rapidly improves with increasing qubit anharmonicity, show that the dominant intrinsic error is caused by a nonadiabatic excitation of the bus $|2\rangle$ state that occurs during the switching the qubit frequency, derive a fidelity estimator based on that mechanism, analyze pulse shape errors, and simulate qubit-qubit {\sf CZ} gates mediated by the bus. Finally, we address the interesting problem of {\it system optimization}, by using gate and idling error estimates to deduce optimal values of resonator frequencies and couplings. 

\section{SUMMARY OF RESULTS}

We find that very high intrinsic fidelites---in the range of 99.9\% to 99.99\% and with corresponding total gate times in the range of  17 to $23 \, {\rm ns}$---can indeed be obtained with a four-parameter gate. Two control parameters are pulse-shape parameters and two are auxiliary local $z$ rotation angles. We emphasize that only low-frequency pulses are required, and that the number of control parameters does not depend on the number of qubits in the QVN device. The results quoted above assume four phase or transmon qubits with $300 \, {\rm MHz}$ anharmonicity; other values of anharmonicity are considered below. The {\sf CZ} gate referred to here is between qubit $q_1$ and the bus (see Fig.~\ref{fig:rezqu}), not between a pair of qubits as is usually considered. 

The two-parameter low frequency pulse profile we use throughout this work is
\begin{widetext}
\BEq\label{pulseEq}
\epsilon(t) = \omega_{\rm off}+\frac{\omega_{\rm on}-\omega_{\rm off}}{2}
\left[ {\rm Erf}\left( \frac{t- \frac{1}{2} t_{\rm ramp}}{ \sqrt{2} \sigma  }\right) - {\rm Erf}\left( \frac{t-t_{\rm gate}
+\frac{1}{2} t_{\rm ramp}}{ \sqrt{2} \sigma  }\right)\right],
\EEq
an example of which is shown in Fig.~\ref{fig:pulse}. Here $\epsilon$ is the qubit frequency, $\omega_{\rm off}$ and $\omega_{\rm on}$ are the frequencies off and near resonance (with the bus), and the pulse switching time is determined by $\sigma$, the standard deviation of the Gaussians inside (\ref{pulseEq}). The value of $t_{\rm ramp}$ determines how the pulse is truncated at $t=0$ and $t_{\rm gate}$ as explained in Sec.~\ref{pulse shape section}; throughout this work we assume that
\begin{equation}
t_{\rm ramp} = 4 \sqrt{2} \sigma.
\label{tramp specification}
\end{equation}
The relation (\ref{tramp specification}) allows the switching time to be alternatively characterized by $t_{\rm ramp}$, which, as Fig.~\ref{fig:pulse} illustrates, is a measure of the width of the ramps. The variable $t_{\rm gate}$ is the total execution time of the gate excluding $z$ rotations. The two control parameters $\omega_{\rm on}$ and 
\begin{equation}
t_{\rm on} \equiv t_{\rm gate} - t_{\rm ramp}
\label{ton definition}
\end{equation}
are determined by the numerical optimization procedure described in Sec.~\ref{optimization section}. From (\ref{ton definition}) we infer that $ t_{\rm on}$ is the time interval between the midpoints of the ramps, or the full-width at half-maximum (FWHM) of the pulse. We note that the optimal values of $t_{\rm on}$ are somewhat longer than the value
\begin{equation}
t_{\rm on}^{\rm sudden} \equiv \frac{\pi}{\sqrt{2}g_{\rm b}}
\label{ton sudden definition}
\end{equation}
that applies in the sudden, $\sigma \rightarrow 0$ limit. In addition to $\omega_{\rm on}$ and $t_{\rm on}$, two auxiliary local $z$ rotations---on the qubit and resonator---are used to implement the {\sf CZ} gate. As we explain below, adjusting the two control parameters $\omega_{\rm on}$ and $t_{\rm on}$ zeros the population left in the non-computational qubit $|2\rangle$ state after the gate and (along with the auxiliary $z$ rotations) sets the controlled phase equal to $-1$. The pulse shape (\ref{pulseEq}) describes a rectangular current or voltage pulse sent to the qubit frequency bias through a Gaussian filter of width $\sigma$, and is believed to be an accurate (although not exact) representation of the actual profile seen by the qubits in Ref.~\cite{MariantoniSci11}.

\begin{table}[htb]
\centering
\caption{\label{fidelity table} Optimal state-averaged process fidelity $F_{\rm ave}$ for the Strauch {\sf CZ} gate between qubit $q_1$ and the bus, in the ${\rm QVN}_4$ processor of Fig.~\ref{fig:rezqu}. No decoherence or noise is included here. Specifications for 99.9\% and 99.99\% gates are provided for three values of qubit anharmonicity $\eta$. The parameters $t_{\rm ramp}$ and $\sigma$ characterize the pulse switching time, and $t_{\rm gate}$ is the total gate time excluding auxiliary $z$ rotations. $F_{|11\rangle}$ is the minimum fidelity. Data after double vertical lines give the nonadiabatic switching error and minimum fidelity estimates; these quantities are defined and discussed in Sec.~\ref{fidelity estimator section}. }
\begin{tabular}{|c|c|c|c|c|c|c|c|c|c||c|c|c|}
\hline  
$\eta/2\pi$ & $g_{\rm b}/2\pi$ & $g_{\rm m}/2 \pi$ & $t_{\rm on}^{\rm sudden}$ & $t_{\rm ramp}$ & $\sigma$ & $t_{\rm on}$ & $t_{\rm gate}$ & $F_{\rm ave}$ & $F_{|11\rangle}$ & $|A|^2$ & $p_{\rm sw}$ & $F_{|11\rangle}^{\rm (est)}$  \\
\hline  
$200 \, {\rm MHz}$  &  $30 \, {\rm MHz}$  & $100 \, {\rm MHz}$ & $11.8 \, {\rm ns}$ & $11 \, {\rm ns}$ & $1.94 \, {\rm ns}$ & $15.8 \, {\rm ns}$ &  $26.8 \, {\rm ns}$ & 99.901\% & 99.613\%    &  $2.1 \! \times \! 10^{-2}$  &  $1.5 \! \times \! 10^{-3}$  &  99.692\%  \\ 
 &    &  & & $16 \, {\rm ns}$ & $2.83 \, {\rm ns}$ & $18.3 \, {\rm ns}$ &  $34.3 \, {\rm ns}$ & 99.992\% &  99.975\%   &  $2.8 \! \times \! 10^{-3}$   &  $2.0 \! \times \! 10^{-4}$  &  99.960\% \\ 
 \hline  
$300 \, {\rm MHz}$  &  $45 \, {\rm MHz}$  & $100 \, {\rm MHz}$ &$7.9 \, {\rm ns}$ & $7 \, {\rm ns}$ & $1.24 \, {\rm ns}$ & $9.9 \, {\rm ns}$ &  $16.9 \, {\rm ns}$ & 99.928\% &   99.714\%  &  $1.7 \! \times \! 10^{-2}$  &  $1.2 \! \times \! 10^{-3}$  &  99.761\%  \\ 
 &    &  & &  $11 \, {\rm ns}$ & $1.94  \, {\rm ns}$ & $11.8 \, {\rm ns}$ &  $22.8 \, {\rm ns}$ & 99.995\% &  99.979\%  &  $9.9 \! \times \! 10^{-4}$  &  $7.2 \! \times \! 10^{-5}$  &  99.986\% \\ 
 \hline  
$400 \, {\rm MHz}$  &  $60 \, {\rm MHz}$  & $100 \, {\rm MHz}$ &$5.9 \, {\rm ns}$ & $5 \, {\rm ns}$ & $0.88 \, {\rm ns}$ & $7.0 \, {\rm ns}$ &  $12.0 \, {\rm ns}$ & 99.950\% & $ 99.804\%$  &  $1.4 \! \times \! 10^{-2}$  &  $1.0 \! \times \! 10^{-3}$  &  99.799\%  \\ 
 &    &  & & $7 \, {\rm ns}$ & $1.24 \, {\rm ns}$ & $7.8 \, {\rm ns}$ &  $14.8 \, {\rm ns}$ & 99.991\% & 99.966\%  &  $2.1 \! \times \! 10^{-3}$   &  $1.5 \! \times \! 10^{-4}$   &  99.970\%  \\ 
 \hline
\end{tabular}
\end{table}
\end{widetext}

Our main results are given in Table \ref{fidelity table}. Here $\eta$ is the qubit anharmonicity. The $200 \, {\rm MHz}$ results apply to the phase qubits of Ref.~\cite{MariantoniSci11}, while the larger anharmonicities might be relevant for future implementations with transmons. The bus couplings $g_{\rm b}$ are determined by the ``$g$ optimization" procedure described in Sec.~\ref{system optimization section}, which leads to the simple formula
\begin{equation}
\frac{g_{\rm b}}{\eta}=0.15,
\label{optimal gb formula}
\end{equation}
for the (approximately) optimal bus coupling. Let $``{\rm QVN}_n \! "$ refer to a quantum von Neumann processor with $n$ qubits coupled to $n$ memory resonators and a bus; the Hamiltonian for such a device is discussed in Sec.~\ref{sec:model}. As indicated in Table \ref{fidelity table}, the memory resonators are always strongly coupled to allow for fast (less than $5 \, {\rm ns}$) {\sf MOVE} operations in and out of memory. {\sf CZ} fidelities well above 99.99\% are also obtainable (see Sec.~\ref{5 nines section}). Table \ref{fidelity table} shows that the time required for a qubit-bus {\sf CZ} gate with fixed intrinsic fidelity is  inversely proportional to the qubit anharmonicity, namely
\begin{equation}
t_{\rm gate}^{\rm (99.9\%)} \approx \frac{5.2}{\eta/2\pi} \ \ \ \ {\rm and} \ \ \ \
t_{\rm gate}^{\rm (99.99\%)} \approx \frac{6.7}{\eta/2\pi} .
\end{equation}
These expressions indicate that ${\sf CZ}$ gates with very high intrinsic fidelity can be implemented in about $20 \, {\rm ns}$ with existing superconducting qubits, a conclusion which applies not only to ${\rm QVN}_n$ but also to a wide range of similar resonator-based architectures. The intrinsic gate (or process) fidelity $F_{\rm ave}$ is the squared overlap of ideal and realized final states, averaged over initial states (see Sec.~\ref{fidelity definition section}). By {\it intrinsic} we mean that noise and decoherence are not included in the gate simulation. The fidelity estimate is developed in Sec.~\ref{fidelity estimator section}. The results given in Table \ref{fidelity table} apply specifically to the $n \! = \! 4$ processor, but similar results are expected for other (not too large) values of $n$. Two strategies are critical for obtaining this high performance: Separating two control parameters in the form of auxiliary local $z$ rotations, and defining the computational states in terms of {\it interacting} system eigenfunctions. These strategies were used in Ref.~\cite{GaliautdinovPRA12} and are discussed in more detail below. The gate fidelities achievable with a transmon-based QVN device are in line with that required for fault-tolerant quantum computation with topological stabilizer codes \cite{BravyiPre98,RaussendorfPRL07,FowlerPRA09}. Qubit anharmonicity is an important resource that will help us achieve that goal. 

The {\sf CZ} gate of Table \ref{fidelity table} is between qubit $q_1$ and the bus in the ${\rm QVN}_4$ device, and begins (typically) with a superposition of qubit-bus eigenstates, with the other qubits and all memory resonators in their ground states. In Sec.~\ref{other CZ gates section} we also comment on several extensions and variations of this basic qubit-bus {\sf CZ} gate: To begin with, the same gate but with qubit $q_4$ (which has a different memory frequency) is considered in Sec.~\ref{qubit 4 CZ gate section}. In Sec.~\ref{CZ between qubits section} we simulate a {\sf CZ} gate between two {\it qubits} in ${\rm QVN_4}$, starting in the idling configuration where the qubits are empty and all data is stored in memory. In this case the qubit-bus {\sf CZ} gate of Table \ref{fidelity table} is supplemented with {\sf MOVE} gates to effect a {\sf CZ} between qubits. And in Sec.~\ref{directly coupled CZ section} we discuss the {\sf CZ} implemented between a pair of {\it directly} coupled anharmonic qubits, instead of a qubit and resonator. This is the system originally considered by Strauch {\it et al.}~\cite{StrauchPRL03}.

\section{{\sf CZ} GATE DESIGN}
\label{CZ gate design section}

In this section we discuss the qubit-bus {\sf CZ} gate design problem.

\subsection{QVN model}
\label{sec:model}

The ${\rm QVN}_n$ processor consists of $n$ superconducting qubits \cite{YouPT05,SchoelkopfNat08,ClarkeNat08} capacitively coupled \cite{JohnsonPRB03,BlaisPRL03,StrauchPRL03} to $n$ memory resonators and to a common bus resonator. Here we assume parameters appropriate either for phase qubits \cite{MartinisQIP09} or transmon qubits \cite{KochPRA07,HouckQIP09} with tunable transition frequencies. We write the qubit angular frequencies as $\epsilon_i$,  with $i=1,\cdots,n$. These are the only controllable parameters in the QVN Hamiltonian (in contrast with Refs.~\cite{YouPRB05,RigettiPRL05,LiuPRL06,ParaoanuPRB06,AshhabPRB06,GrajcarPRB06,AshhabPRB07,LiuPRB07,RigettiPRB10,deGrootNatPhys10,ChowPRL11,deGrootNJP12,ChowPRL12} we do not make use of microwave pulses). The memory frequencies are written as $\omega_{{\rm m}i}$, and the bus frequency is $\omega_{\rm b}$. The (bare) frequencies of all resonators are assumed here to be fixed.

\BW

Because we are interested in very high fidelities, a realistic model is required. However, we have shown (in unpublished work) that the {\sf CZ} performance is extremely robust with respect to the model details, so we only report results for a simplified Hamiltonian; the approximations used are discussed below. For the qubit-bus {\sf CZ} simulations, the Hilbert space  is truncated to allow for up to three excitations. The {\sf CZ} gate naively involves no more than two excitations, so to properly account for leakage we include up to three. Therefore, four-level qubits and resonators (which include the $|3\rangle$ states) are required in the model. The QVN Hamiltonian we use is
\BEq
H = \sum_{i=1}^n \left[\begin{pmatrix}
 0 & 0 & 0 & 0 \\
 0 & \epsilon_{i} & 0 & 0 \\
 0 & 0 & 2\epsilon_{i} -\eta & 0 \\
 0 & 0 & 0 & 3\epsilon_{i}-\eta^\prime \\
 \end{pmatrix}_{\! \! \! \!  {\rm q}i} + \begin{pmatrix}
 0 & 0 & 0 & 0 \\
 0 & \omega_{{\rm m}i} & 0 & 0 \\
 0 & 0 & 2\omega_{{\rm m}i} & 0 \\
 0 & 0 & 0 & 3\omega_{{\rm m}i} \\
 \end{pmatrix}_{\! \! \! \!  {\rm m}i} 
 + g_{\rm m} \, Y_{{\rm q}i} \otimes Y_{{\rm m}i} 
  + g_{\rm b} \, Y_{{\rm q}i} \otimes Y_{{\rm b}} 
 \right]+\begin{pmatrix}
 0 & 0 & 0 & 0 \\
 0 & \omega_{\rm b} & 0 & 0 \\
 0 & 0 & 2\omega_{\rm b} & 0 \\
 0 & 0 & 0 & 3\omega_{\rm b} \\
 \end{pmatrix}_{\! \! \! \!  {\rm b}} ,
 \label{QVN hamiltonian}
\EEq
\EW
excluding single-qubit terms for microwave pulses that are not used in this work. Here $\eta$ and $\eta^\prime $ are qubit anharmonic detuning frequencies, $g_{\rm m}$ and $g_{\rm b}$ are the qubit-memory and qubit-bus interation strengths, and
\BEq
Y \equiv \begin{pmatrix}
 0 & -i & 0 & 0 \\
 i & 0 & -\sqrt{2}i & 0 \\
 0 & \sqrt{2}i & 0 & -\sqrt{3}i \\
 0 & 0 & \sqrt{3}i & 0 
 \label{Y matrix}
 \end{pmatrix}.
\EEq
The matrices in (\ref{QVN hamiltonian}) act nontrivially in the spaces indicated by their subscripts, and as the identity otherwise. The matrix $Y$ results from a harmonic oscillator approximation for the qubit eigenfunctions. Factors of $\hbar$ are suppressed throughout this paper. 

The main approximations leading to (\ref{QVN hamiltonian}) are the neglect of the $\epsilon$-dependence of the interaction strengths $g_{\rm m}$ and $g_{\rm b}$, and the neglect of a small direct coupling between the memories and bus \cite{directCouplingNote}. We have verified that including these does not change the main conclusions of this work. The $\epsilon$-dependence of the anharmonicities $\eta$ and $\eta^\prime$, and small anharmonic corrections to the interaction terms in (\ref{QVN hamiltonian}), are also neglected. 

The parameter values we use in our simulations are provided in Table \ref{parameter table}. We assume $\eta'=3\eta$, which is appropriate for qubic anharmonicity. As discussed in Sec.~\ref{system optimization section}, the value of the bus coupling $g_{\rm b}$ is chosen to give the shortest {\sf CZ} gate time (for a range of fidelities). The choice of resonator frequencies is also discussed in Sec.~\ref{system optimization section}. We simulate $n=4$ qubits. The fidelities quoted in this paper are numerically exact for the model (\ref{QVN hamiltonian}); the rotating-wave approximation is not used.

\begin{table}[htb]
\centering
\caption{\label{parameter table} Device parameters used in this work.}
\begin{tabular}{|c|c|}
\hline
quantity & value \\
\hline 
empty qubit parking frequency $\omega_{\rm park}/2 \pi$  & $10.0 \, {\rm GHz}$ \\
\hline 
memory resonator $m_1$ frequency $\omega_{{\rm m}1}/2 \pi$  & $8.3 \, {\rm GHz}$ \\
memory resonator $m_2$ frequency $\omega_{{\rm m}2}/2 \pi$  & $8.2 \, {\rm GHz}$ \\
memory resonator $m_3$ frequency $\omega_{{\rm m}3}/2 \pi$  & $8.1 \, {\rm GHz}$ \\
memory resonator $m_4$ frequency $\omega_{{\rm m}4}/2 \pi$  & $8.0 \, {\rm GHz}$ \\
\hline
initial detuned qubit frequency $\omega_{\rm off}/2 \pi$  & $7.5 \, {\rm GHz}$ \\
\hline 
bus resonator frequency $\omega_{\rm b}/2 \pi$  & $6.5 \, {\rm GHz}$ \\
\hline 
qubit-memory coupling strength $g_{\rm m}/2 \pi$  & $100 \, {\rm MHz}$ \\
\hline 
qubit-bus coupling strength $g_{\rm b}/2 \pi$  & $30-60 \, {\rm MHz}$ \\
\hline 
qubit anharmonicity $\eta/2 \pi $   & $200-400 \, {\rm MHz}$  \\
\hline
\end{tabular}
\end{table}

Although the {\sf CZ} and {\sf MOVE} gates considered here do not involve microwave pulses, the single-qubit gates are assumed to be implemented with microwaves in the usual manner at the qubit frequency $\omega_{\rm off}$. This frequency is also used to define an experimental ``rotating" reference frame or local clock for each qubit: All qubit frequencies are defined relative to $\omega_{\rm off}$ \cite{MariantoniSci11}. This is  discussed below in Sec.~\ref{eigenstate basis section}.

\subsection{Strauch {\sf CZ} gate}
\label{Strauch CZ gate section}

In this section we give a detailed description of the {\sf CZ} gate introduced by Strauch {\it et al.}~\cite{StrauchPRL03}. In particular, we explain the specific roles played by the two pulse-shape parameters $t_{\rm on}$ and $\omega_{\rm on}$, and by the two auxiliary $z$ rotation angles $\gamma_1$ and $\gamma_2$. To accomplish this we introduce several approximations that allow for an analytic treatment of the {\sf CZ} gate dynamics. 

\begin{figure}[htb]
\centering
\includegraphics[angle=0,width=1.00\linewidth]{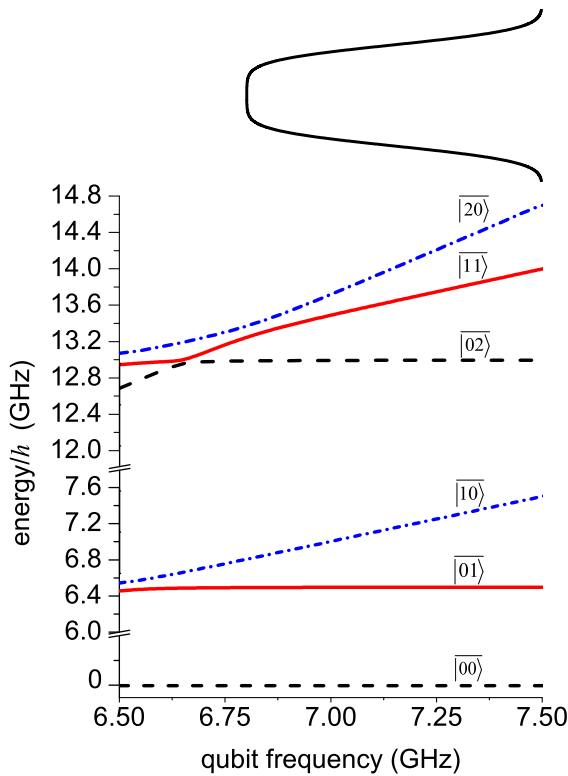}
\caption{(Color online) Energies of eigenstates $\overline{\ket{qb}}$ of a single qubit $q$ coupled to a resonator bus $b$. Here $\omega_{\rm b}/2\pi=6.5 \, {\rm GHz}$, $\eta/2\pi=300 \, {\rm MHz}$, and $g_{\rm b}/2 \pi = 45 \, {\rm MHz}$. The time dependence of the qubit frequency during a {\sf CZ} gate (solid black curve) is indicated at the top of the figure.}
\label{fig:energylevels}
\end{figure}

First, we consider a truncated model consisting of a single superconducting qubit with frequency $\epsilon$ and anharmonic detuning $\eta$, capacitively coupled to a bus resonator with frequency $\omega_{\rm b}$,
\begin{equation}
H = \begin{pmatrix}
 0 & 0 & 0 \\
 0 & \epsilon & 0 \\
 0 & 0 & 2\epsilon -\eta \\
 \end{pmatrix}_{\! \! \! \!  {\rm q}} 
  + \begin{pmatrix}
 0 & 0 & 0 \\
 0 & \omega_{\rm b} & 0 \\
 0 & 0 & 2\omega_{\rm b}  \\
 \end{pmatrix}_{\! \! \! \!  {\rm b}} 
  + g_{\rm b} \, Y_{\rm q} \otimes Y_{\rm b}.
\label{qb hamiltonian}
\end{equation}
In this case $Y$ reduces to
\BEq
Y = \begin{pmatrix}
 0 & -i & 0 \\
 i & 0 & -\sqrt{2}i \\
 0 & \sqrt{2}i & 0  \\
 \end{pmatrix} \! .
\EEq
This Hamiltonian is written in the basis of {\it bare} eigenstates, which are the system eigenfunctions when the qubit and resonator are uncoupled. We write these bare states as $\ket{qb}$, with $q,b \in \lbrace 0,1,2\rbrace.$ The energies of the interacting eigenstates, which we write with an overline as $\overline{\ket{qb}}$, are plotted in Fig.~\ref{fig:energylevels} as a function of $\epsilon/2 \pi$ for the case of $\omega_{\rm b}/2\pi=6.5 \, {\rm GHz}$, $\eta/2\pi=300 \, {\rm MHz}$, and $g_{\rm b}/2 \pi = 45 \, {\rm MHz}$. The interacting eigenstates are labeled such that $\overline{\ket{qb}}$ is perturbatively connected to $\ket{qb}$ when $\epsilon \gg \omega_{\rm b}$.

Second, we assume a short switching time and ignore the dynamical phases acquired during the ramps. As we will see below, this approximation is valid when $g_{\rm b} \ll \eta$, so that the switching can be made {\it sudden} with respect to the coupling $g_{\rm b}$, but still adiabatic with respect to the anharmonicity $\eta$.

The {\sf CZ} gate of Strauch {\it et al.}~\cite{StrauchPRL03}, adapted to the qubit-resonator system, works by using the anticrossing of the $\overline{\ket{11}}$ channel with the auxiliary state $\overline{\ket{20}}$. In terms of the pulse parameters defined in (\ref{pulseEq}), the qubit-resonator state is prepared at a qubit frequency $\epsilon=\omega_{\rm off}$, and the frequency is then switched to $\epsilon=\omega_{\rm on}$ for a FWHM time duration $t_{\rm on}$. In the simplified model considered in this section,
\begin{equation}
\omega_{\rm on} = \omega_{\rm b} + \eta,
\label{omega_on formula for simplified model}
\end{equation}
and 
\begin{equation}
t_{\rm on}= \frac{\pi}{\sqrt{2} g_{\rm b}}.
\label{t_on formula for simplified model}
\end{equation}
Equation (\ref{omega_on formula for simplified model}) gives the qubit frequency for which the bare state $|11\rangle$ is degenerate with $|20\rangle$, and is at a frequency $\eta$ above the usual resonance condition. Equation (\ref{t_on formula for simplified model}) is the sudden-limit value defined in (\ref{ton sudden definition}) and derived below. The qubit frequency is then returned to the detuned value $\omega_{\rm off}$. The complete pulse profile is also shown in Fig.~\ref{fig:energylevels} (solid black curve) for the case of $\omega_{\rm on}/2\pi=6.8 \, {\rm GHz}$ and $\omega_{\rm off}/2\pi=7.5 \, {\rm GHz}$. 

Let's follow the evolution resulting from an initial (normalized) qubit-resonator state
\begin{equation}
a_{00} \overline{\big| 00 \big\rangle} + a_{01} \overline{\big| 01 \big\rangle} + a_{10} \overline{\big| 10 \big\rangle}  + a_{11} \overline{\big| 11 \big\rangle}.
\end{equation}
Because the $\overline{|00\rangle}$ channel is very well separated from the others, the $\overline{|00\rangle}$ component will only acquire a dynamical phase factor
\begin{equation}
e^{-i E_{00} t_{\rm gate}},
\label{00 phase factor}
\end{equation}
where $E_{00}$ is the energy of the $\overline{\big| 00 \big\rangle}$ eigenstate. Without loss of generality we can shift the entire spectrum so that $E_{00}=0$ [as in (\ref{qb hamiltonian})] and the phase factor (\ref{00 phase factor}) becomes unity. This freedom results from the fact that any unitary gate operation only needs to be defined up to an overall multiplicative phase factor. With this phase convention the {\sf CZ} gate acts as the identity on this component, so we have the map
\begin{equation}
\overline{\big| 00 \big\rangle} \rightarrow \overline{\big| 00 \big\rangle} .
\label{CZ evolution 00 component}
\end{equation}

The $\overline{|01\rangle}$ component will mostly return to $\overline{|01\rangle}$, also with an acquired phase, but a small component will be left in $\overline{|10\rangle}$ due to the nonadiabatic excitation of that channel, which is only separated in energy from $\overline{|01\rangle}$ by about $\eta$ when $\epsilon = \omega_{\rm b} + \eta$. The $\overline{|10\rangle}$ component similarly suffers from a small noniadabatic coupling to $\overline{|01\rangle}$. As we will explain below, these nonadiabatic errors are exponentially suppressed when the functional form of $\epsilon(t)$ is properly designed. Then we have
\begin{equation}
\overline{\big| 01 \big\rangle} \rightarrow e^{-i \alpha} \sqrt{1-\mathbb{E}_1} \ \overline{\big| 01 \big\rangle} + e^{-i \alpha^\prime} \sqrt{\mathbb{E}_1} \ \overline{\big| 10 \big\rangle} 
\label{CZ evolution 01 component general}
\end{equation}
and
\begin{equation}
\overline{\big| 10 \big\rangle} \rightarrow e^{-i \beta} \sqrt{1-\mathbb{E}_1} \ \overline{\big| 10 \big\rangle} + e^{-i \beta^\prime}  \sqrt{\mathbb{E}_1} \ \overline{\big| 01 \big\rangle}, 
\label{CZ evolution 10 component general}
\end{equation}
where $\mathbb{E}_1$ is a small nonadiabatic population error (below we refer to $\mathbb{E}_1$ as a {\it switching} error). In the $\mathbb{E}_1 \rightarrow 0$ limit, $\alpha$ and $\beta$ are dynamical phases given by
\begin{eqnarray}
\alpha &=& \int_0^{t_{\rm gate}} \! \! E_{01} \, dt \approx \bigg(\omega_{\rm b} - \frac{g_{\rm b}^2}{\eta} \bigg) t_{\rm on}, 
\label{dynamical phase alpha} \\
\beta &=& \int_0^{t_{\rm gate}} \! \! E_{10} \, dt \approx \bigg(\omega_{\rm b} + \eta + \frac{g_{\rm b}^2}{\eta} \bigg) t_{\rm on}, 
\label{dynamical phase beta}
\end{eqnarray}
where the second approximate quantities neglect phase accumulation during the ramps and use perturbative expressions for the energies $E_{01}$ and $E_{10}$ when $\epsilon=\omega_{\rm b} + \eta$. The expressions (\ref{CZ evolution 01 component general}) and (\ref{CZ evolution 10 component general}) neglect an extremely small leakage out of the $\lbrace \overline{|01\rangle} , \overline{|10\rangle} \rbrace$ subspace. Neglecting this leakage, the evolution in the $\lbrace \overline{|01\rangle} , \overline{|10\rangle} \rbrace$ subspace is unitary, leading to the phase condition 
\begin{equation}
e^{i( \alpha- \beta^\prime)} + e^{i( \alpha^\prime- \beta)} = 0.
\label{subspace unitarity condition}
\end{equation}
Using (\ref{subspace unitarity condition}) to eliminate $\beta^\prime$ leads to
\begin{eqnarray}
\overline{\big| 01 \big\rangle} &\rightarrow& e^{-i \alpha} \sqrt{1-\mathbb{E}_1} \ \overline{\big| 01 \big\rangle} + e^{-i (\beta+\phi)} \sqrt{\mathbb{E}_1} \ \overline{\big| 10 \big\rangle} ,
\label{CZ evolution 01 component} \\
\overline{\big| 10 \big\rangle} &\rightarrow& e^{-i \beta} \sqrt{1-\mathbb{E}_1} \ \overline{\big| 10 \big\rangle} - e^{-i(\alpha-\phi)}  \sqrt{\mathbb{E}_1} \ \overline{\big| 01 \big\rangle}, 
\label{CZ evolution 10 component}
\end{eqnarray}
where $\phi \equiv \alpha^\prime - \beta$. The evolution of the eigentates $\overline{|01\rangle}$ and $\overline{|10\rangle}$ is therefore characterized by the cross-excitation probability $\mathbb{E}_1$ and three phase angles $\alpha$, $\beta$, and $\phi$.
   
Now we consider the $\overline{|11\rangle}$ component. The $\overline{|11\rangle}$ channel couples strongly with the $\overline{|20\rangle}$ channel, as well as weakly with $\overline{|02\rangle}$. The simplest way to understand the dynamics of the $\overline{|11\rangle}$ component is to use two different representations to describe these two effects. We will describe strong interaction with $\overline{|20\rangle}$ in the bare basis and the weak, nonadiabatic coupling with $\overline{|02\rangle}$ in the eigenstate basis. Suppose we begin with the qubit strongly detuned from the bus, so that $\overline{|11\rangle} \approx |11\rangle$ (the detuned interacting eigenstate is well approximated by the bare $|11\rangle$ state). Then we quickly switch $\epsilon$ from $\omega_{\rm off}$ to $\omega_{\rm b}+ \eta.$ By ``quickly" we mean that we strongly mix with the $\overline{|20\rangle}$ channel. The interaction with $\overline{|02\rangle}$ is always weak, even in the sudden limit. This asymmetric excitation is possible because $\overline{|20\rangle}$ is protected (separated in energy from $\overline{|11\rangle}$) by an energy gap $2 \sqrt{2} g_{\rm b}$, whereas $\overline{|02\rangle}$ is protected by a much larger gap of $\eta - \sqrt{2} g_{\rm b}$ (this expression accounts for level repulsion from $\overline{|20\rangle}$, and we have assumed that $g_{\rm b} \ll \eta$). We can informally say that the desired switching is nonadiabatic with respect to the energy scale $g_{\rm b}$, but is adiabatic with respect to $\eta$ \cite{StrauchPRL03}.

Focusing first on the strong coupling to $\overline{|20\rangle}$, the suddenly switched $|11\rangle$ state is no longer an eigenstate when $\epsilon = \omega_{\rm on}$, as the relevant eigenfunctions at this setting are
\begin{equation}
\overline{|11\rangle} = \frac{|11\rangle - |20\rangle}{\sqrt{2}} \ \ \ {\rm and} \ \ \
\overline{|20\rangle} = \frac{|11\rangle + |20\rangle}{\sqrt{2}}.
\end{equation}
The nonstationary state
\begin{equation}
|11\rangle = \frac{\overline{|11\rangle}+\overline{|20\rangle}}{\sqrt{2}} 
\label{tuned nonstationary state}
\end{equation}
therefore rotates in the $\lbrace |11\rangle , |20\rangle \rbrace$ subspace, and after a time duration $t$ becomes
\begin{eqnarray}
|\psi\rangle = e^{-i E_{11}t} &\bigg[& \frac{\overline{|11\rangle} + e^{-i \Delta \! E  t} \overline{|20\rangle}}{\sqrt{2}} \bigg] \\
= e^{-i E_{11}t} &\bigg[& \bigg(\frac{1+e^{-i \Delta \! E t}}{2}  \bigg) \big|11\big\rangle \nonumber \\
&-& \bigg(  \frac{1-e^{-i \Delta \! E t}}{2} \bigg)  \big|20\big\rangle\bigg], 
\label{nonstationary 11 state}
\end{eqnarray}
where
\begin{equation}
\Delta E \equiv E_{20}-E_{11} = 2 \sqrt{2} g_{\rm b}.
\end{equation}
Holding $\epsilon$ fixed at $\omega_{\rm b} + \eta$ for a FWHM time (\ref{t_on formula for simplified model}), corresponding to a $2\pi$ rotation, (\ref{nonstationary 11 state}) becomes
\begin{equation}
|\psi\rangle = e^{-i E_{11}t_{\rm on}} |11\rangle .
\end{equation}
When $\epsilon=\omega_{\rm b} + \eta$, the energy of eigenstate $\overline{|11\rangle}$ is
\begin{equation}
E_{11} = 2 \omega_{\rm b} + \eta - \sqrt{2} g_{\rm b}.
\end{equation}
After detuning quickly we therefore obtain
\begin{equation}
\overline{\big|11\big\rangle} \rightarrow - \exp\bigg[ {-i \bigg ( \pi \frac{2 \omega_{\rm b} +\eta}{\sqrt{2} g_{\rm b}} } \bigg) \bigg] \, \overline{\big|11\big\rangle},
\end{equation}
or, using expressions (\ref{dynamical phase alpha}) and (\ref{dynamical phase beta}),
\begin{equation}
\overline{\big|11\big\rangle} \rightarrow - e^{-i(\alpha+\beta)} \, \overline{\big|11\big\rangle}.
\label{CZ evolution 11 component without switching error}
\end{equation}

The two phase angles $\alpha$ and $\beta$ can be cancelled by the application of independent auxiliary single-qubit $z$ rotations
\begin{equation}
R_z(\gamma) \equiv \exp[{-i (\gamma/2)\sigma^z}] 
\label{z rotation definition}
\end{equation}
to the qubit and bus. Qubit $z$ rotations are implemented by frequency excursions, whereras resonator $z$ rotations are implemented in software (they are compiled into future qubit rotations). Following the pulse sequence that leads to (\ref{CZ evolution 00 component}), (\ref{CZ evolution 01 component}), (\ref{CZ evolution 10 component}), and (\ref{CZ evolution 11 component without switching error}), with the operation
\begin{equation}
R_z(\gamma_1) \otimes R_z(\gamma_2) ,
\end{equation}
where
\begin{equation}
\gamma_1 = -\beta \ \ \ {\rm and} \ \ \ \gamma_2 = -\alpha,
\end{equation}
leads to the map
\begin{eqnarray}
\overline{\big| 00 \big\rangle} &\rightarrow&  \overline{\big| 00 \big\rangle},  \label{CZ evolution 00 component after z rotations} \\
\overline{\big| 01 \big\rangle} &\rightarrow&  \sqrt{1-\mathbb{E}_1} \ \overline{\big| 01 \big\rangle} + e^{-i \phi} \sqrt{\mathbb{E}_1} \ \overline{\big| 10 \big\rangle},  \label{CZ evolution 01 component after z rotations} \\
\overline{\big| 10 \big\rangle} &\rightarrow& \sqrt{1-\mathbb{E}_1} \ \overline{\big| 10 \big\rangle} - e^{i \phi}  \sqrt{\mathbb{E}_1} \ \overline{\big| 01 \big\rangle}, \label{CZ evolution 10 component after z rotations} \\
\overline{\big|11\big\rangle} &\rightarrow& -  \overline{\big|11\big\rangle}, \label{CZ evolution 11 component after z rotations}
\end{eqnarray}
apart from a global phase factor. The use of auxiliary $z$ rotations is discussed further Sec.~\ref{z rotation section}.

The minus sign in (\ref{CZ evolution 11 component after z rotations}) is the key to the Strauch {\sf CZ} gate. However, as mentioned above, the analysis leading to (\ref{CZ evolution 11 component after z rotations}) neglected a weak nonadiabatic excitation of the $\overline{|02\rangle}$ channel caused by the switching of $\epsilon$. Including this effect in (\ref{CZ evolution 11 component after z rotations}) leads to the modification
\begin{eqnarray}
\overline{\big|11\big\rangle} \rightarrow &-& \sqrt{1-\mathbb{E}_2} \ \overline{\big|11\big\rangle} \nonumber \\
&+& {\rm phase \ factor} \times \sqrt{\mathbb{E}_2} \ \overline{\big|02\big\rangle},
\label{CZ evolution 11 component with switching error}
\end{eqnarray}
where $\mathbb{E}_2$ is another switching error. Both $\mathbb{E}_1$ and $\mathbb{E}_2$ vanish exponentially with $\sigma$ (or $t_{\rm ramp}$), and for the regimes studied in this work $\mathbb{E}_2$ is the dominant source of intrinsic gate fidelity loss. We note that the analysis leading to (\ref{CZ evolution 11 component with switching error}) assumed implementation of the {\it ideal} values [(\ref{omega_on formula for simplified model}) and (\ref{t_on formula for simplified model})] of $\omega_{\rm on}$ and $t_{\rm on}$. Errors in these two control parameters, which we refer to as pulse shape errors and study in Sec.~\ref{pulse shape errors section}, lead instead to
\begin{eqnarray}
\overline{\big|11\big\rangle} \rightarrow &-& e^{i\delta} \sqrt{1-\mathbb{E}_2 - \mathbb{E}_\theta} \ \overline{\big|11\big\rangle} \nonumber \\
&+& {\rm phase \ factor} \times \sqrt{\mathbb{E}_2} \ \overline{\big|02\big\rangle} \nonumber \\
&+& {\rm phase \ factor} \times \sqrt{\mathbb{E}_\theta} \ \overline{\big|20\big\rangle},
\label{CZ evolution 11 component with switching and pulse shape errors}
\end{eqnarray}
where the controlled-phase error angle $\delta$ and rotation error $\mathbb{E}_\theta$ depend on the errors in $\omega_{\rm on}$ and $t_{\rm on}$, respectively.

Finally, it is also interesting to consider the fully adiabatic limit of the Strauch {\sf CZ} gate. By this we mean that the switching is adiabatic with respect to both $g_{\rm b}$ and $\eta$. For the gate time to be competitive with the nonadiabatic gate of Table \ref{fidelity table}, a larger coupling $g_{\rm b}$ is required, which might lead to significant higher-order and cross-coupling errors in a multi-qubit device, but in the fully adiabatic limit only {\it one} pulse control parameter---either $\omega_{\rm on}$ or $t_{\rm on}$---needs to be optimized (two $z$ rotations are still required). This is because adiabaticity now assures that the $\overline{|11\rangle}$ population is preserved (apart from exponentially small switching errors), taking over the role previously played by $t_{\rm on}$, and a single pulse shape parameter  is sufficient to specify the controlled phase. A highly adiabatic {\sf CZ} gate was demonstrated in Ref.~\cite{DiCarloNat09}.

There are a few important differences between the Strauch {\sf CZ} gate applied to a pair of directly coupled qubits (as in Ref.~\cite{StrauchPRL03}) and to the qubit-bus system considered here. These differences result from the harmonic spectrum of the resonator in the latter case and are discussed below in Sec.~\ref{directly coupled CZ section}.

\subsection{Eigenstate basis}
\label{eigenstate basis section}

The Hamiltonian (\ref{QVN hamiltonian}) is written in the usual {\it bare} basis of uncoupled system eigenstates, but information processing itself is best performed in the basis of {\it interacting} eigenfunctions of $H_{\rm idle}$, where $H_{\rm idle}$ is given by (\ref{QVN hamiltonian}) with the qubits in a dispersive idling configuration \cite{GaliautdinovPRA12}. This choice of computational basis assures that idling qubits suffer no population change in the decoherence-free limit, and evolve in phase in a way that can be almost exactly compensated for by an appropriate choice of only $n$ rotating frames or local clocks, one for each qubit \cite{GaliautdinovPRA12}. Here we briefly review this important concept.

In principal, any complete orthonormal basis of the physical Hilbert space that can be appropriately prepared, unitarily transformed, and measured---essentially, any basis where one can implement the DiVincenzo criteria \cite{DiVincenzoArxiv0002077}---is a valid basis on which to run a quantum computation. Defining the computational states to be interacting system eigenfunctions gives them the simplifying property that the time evolution can be decomposed into a sequence of gates, between which (almost) no evolution occurs. In other words, idling between gates generates the identity operation. This property, which is implicitly assumed in the standard circuit model of quantum computation, could be realized in an architecture where the Hamiltonian $H$ can be completely switched off between gates. However, it is not possible to set $H=0$ in the QVN architecture; nor can $H$ itself be made negligibly small between gates. Therefore, nonstationary states such as uncoupled-qubit eigenstates accumulate errors (including population oscillations) between gates unless a correction protocol such as dynamical decoupling \cite{ViolaPRA98} is used. By defining computational states in terms of interacting system eigenfunctions $\lbrace \overline{|\psi \rangle} \rbrace$ at some predefined dispersive idling configuration (qubit frequencies), the only evolution occuring during an idle from time $t_1$ to $t_2$ is a pure phase evolution,
\begin{equation}
\overline{|\psi(t_1)\rangle} \rightarrow \overline{|\psi(t_2)\rangle} = e^{-i E  \, (t_2-t_1)} \overline{|\psi(t_1)\rangle},
\label{eigenfunction phase evolution}
\end{equation}
where $E$ is the exact energy eigenvalue (and we neglect decoherence). Furthermore, it is possible to {\it compensate} for---or effectively remove---the pure phase evolution in (\ref{eigenfunction phase evolution}) by applying phase shifts (after the idle period) to each eigenfunction to cancel the $e^{-i E (t_2-t_1)}$ phase factors; doing so would result in the ideal between-gate evolution
\begin{equation}
\overline{|\psi(t_1)\rangle} \rightarrow \overline{|\psi(t_2)\rangle} = \overline{|\psi(t_1)\rangle}.
\label{compensated eigenfunction phase evolution}
\end{equation}
The idling dynamics (\ref{compensated eigenfunction phase evolution}) is evidently equivalent to setting $H=0$ between gates. We will discuss below how the compensating phase shifts are actually implemented in practice. 

This use of interacting system eigenfunctions and compensating phase shifts as described above provides a computational basis that evolves ideally between gates, but such an approach is not scalable; for example, there are $2^{2n+1}$ such computational states in ${\rm QVN}_n$. In Ref.~\cite{GaliautdinovPRA12} an approximate but scalable implementation of this approach was introduced. The idea is that the exact energy $E$ of a computational state in ${\rm QVN}_n$ is, to an extremely good approximation, the sum of uncoupled qubit and resonator frequencies, i.e., essentially noninteracting. This is {\it not} simply a consequence of the dispersive regime energies (eigenvalues of $H_{\rm idle}$), which have non-negligible interaction corrections, but because only a special subset of the eigenfunctions are used for information processing: In the ${\rm QVN}_n$ system we only make use of $H_{\rm idle}$ eigenfunctions in which there are no more than $n$ excitations present, and such that two directly coupled elements---qubits or resonators---are not simultaneously occupied (except during the {\sf CZ} gate). For example, when the data is stored in memory, the residual memory-memory coupling is fourth order in the qubit-resonator coupling $g$ (for simplicity we assume here that $g_{\rm b} \! = \!  g_{\rm m}$). This leads to an eighth-order {\it conditional} frequency shift (order $g^{16}$ idling error) \cite{GaliautdinovPRA12}. Next, suppose an excitation is transferred from memory to a qubit via a {\sf MOVE} gate. Now the dominant frequency shift is sixth order. And when an excitation is in the bus the largest shift is fourth order \cite{GaliautdinovPRA12}. The largest idling error (associated with the phase compensation) is therefore eighth order in $g$ and can be made negligible with proper system design.

The compensating phase shifts could be implemented through additional local $z$ rotations, one for each qubit and resonator. However, these phase shifts evolve in time with very high ($> \! 1 \, {\rm GHz}$) frequency, and it is therefore experimentally more practical to introduce a local clock/rotating frame for each qubit and resonator. This is achieved by introducing a fixed-frequency microwave line for each qubit and resonator, and measuring each qubit and resonator {\it phase} relative to the phase of its reference. By choosing the frequency of the qubit (resonator) reference microwave equal to the idle frequency (resonator frequency), the component frequencies [and therefore the quantity $E$ in (\ref{eigenfunction phase evolution})] are effectively zeroed, and no more than $2n+1$ different reference frequencies or local clocks are required. This procedure corresponds to implementing the experiment in a multi-qubit rotating frame. And, in a further simplification, the local clocks/rotating frames for the resonators are replaced by additional qubit $z$ rotations that are handled in software (i.e., combined with future rotations). Therefore, in practice only $n$ local clocks/rotating frames are needed, one for each qubit.

Because the {\sf CZ} gate simulations reported in Table \ref{fidelity table} are already supplemented with local $z$ rotations, these local clocks/rotating frames do not need to be included in those simulations; we simulate the lab frame. However, they are included in the pulse-shape error simulations reported after (\ref{simplified fidelity loss with uncompensated z rotations}) and in the ${\sf CZ}_{23}$ qubit-qubit gate simulation reported in Sec.~\ref{CZ between qubits section}.

Having motivated the use of interacting system eigenfunctions for computational basis states, it is still necessary to establish that such states can actually be prepared and measured. Because we can assume the processor to initially start in its interacting ground state---a computational basis state---preparation of the other computational states can be viewed as a series of $\pi$ pulses and ${\sf MOVE}$ gates. We expect that such operations on the interacting eigenfunctions can be performed at least as accurately as when applied to bare states. Eigenfunction readout is a more subtle (and model-dependent) question, but the analysis of Ref.~\cite{GaliautdinovPRA12} suggests that interacting-eigenfunction readout is actually better than bare-state readout (in the model considered there).

We also note that the idling configuration and associated eigenstate basis generally changes between consecutive gates (an example is given below in  Sec.~\ref{CZ between qubits section}). In Table \ref{fidelity table}, the idling configuration has qubit $q_1$ at $\omega_{\rm off}$ and the others at $\omega_{\rm park}$. Therefore, our entangling gate design is constrained by the requirement that we start and end in eigenstates of this particular $H_{\rm idle}$.

The discussion above motivating the use of interacting eigenstates is based on their nearly ideal idling dynamics. It is still interesting, then, to consider whether the {\sf CZ} gate can be generated equally well in either (bare or interacting eigenfunction) basis. We find that for the parameter regimes considered here, it is not possble to achieve better than about 99\% fidelity in the bare basis with the same two-parameter pulse profile (it should be possible using more complex pulse shapes). The remaining error is consistent with the size of the perturbative corrections to the bare states in the idling configuration. This exercise emphasizes the importance of performing quantum logic with the system eigenfunctions, which have the built-in protection of {\it adiabiticity} against unwanted transitions. 

One might object to the use of interacting eigenfunctions as a design tool, the exact calculation of which is not scalable. However, approximate dispersive-regime eigenfunctions are efficiently computable. A particularly simple way to do this is to calculate the generator $S$ of the diagonalizing transformation $V=e^{-iS}$ by a power series in $g_{\rm b}$ and $g_{\rm m}$.  At the 99.99\% fidelity level, it is sufficient to calculate $S$ to first order. Writing $H_{\rm idle} = H_0 + \delta H$ leads to the condition $i[S,H_0] + \delta H = 0,$ which is immediately solvable in the bare basis $ |q_1 q_2 \cdots m_1 m_2 \cdots b\rangle.$ Here $q_i, m_i, b \in \lbrace 0,1,2,\dots \rbrace $. Other efficient eigenfunction approximation schemes are also possible. 

In this work we denote the exact or approximate $H_{\rm idle}$ eigenfunction perturbatively connected to the bare state $ | q_1 q_2 \cdots q_n m_1 m_2 \cdots m_n b \rangle$ by
\begin{equation}
\overline{| q_1 q_2 \cdots q_n m_1 m_2 \cdots m_n b \rangle},
\label{QVN eigenstate}
\end{equation}
following the overline notation introduced above. Note that (\ref{QVN eigenstate}) is {\it not} a tensor product of single-qubit/resonator eigenstates as is usually the case.

\subsection{Auxiliary $z$ rotations and {\sf CZ} equivalence class}
\label{z rotation section}

The standard {\sf CZ} gate in the bare two-qubit basis $\lbrace |00\rangle, |01\rangle, |10\rangle, |11\rangle \rbrace$ is
\begin{equation}
{\sf CZ} \equiv \begin{pmatrix}
 1 & 0 & 0 & 0 \\
 0 & 1 & 0 & 0 \\
 0 & 0 & 1 & 0 \\
 0 & 0 & 0 & -1 
 \end{pmatrix} \! .
 \label{CZ definition}
 \end{equation}
However in the QVN processor, local $z$ rotations can be performed quickly and accurately, typically by brief qubit frequency excursions. Thus, we will consider the limit where SU(2) operations of the form $\exp[{-i(\theta/2)\sigma^z}] \! $ can be done on the qubits and bus with negligible error and in a negligible amount of time (fidelity loss resulting from errors in these rotations are discussed in Sec.~\ref{pulse shape errors section}). We therefore want to define our entangling gate modulo these $z$ rotations. We will do this by constructing a {\it local-$z$ equivalence class} for an arbitrary  element (gate) in SU(4), and then specialize to the {\sf CZ} gate.
  
We define two elements $U$ and $U'$ of SU(4) to be equivalent, and write $U' \circeq U$,  if 
\begin{equation}
U'= u_{\rm post} \, U \, u_{\rm pre},
\end{equation} 
where
\begin{widetext}
\begin{equation}
u(\gamma_1,\gamma_2) \equiv R_z(\gamma_1) \otimes R_z(\gamma_2) = 
e^{i(\gamma_1 + \gamma_2)/2}
\begin{pmatrix}
 1 & 0 & 0 & 0 \\
 0 & e^{-i \gamma_2} & 0 & 0 \\
 0 & 0 & e^{-i \gamma_1}  & 0 \\
 0 & 0 & 0 & e^{-i (\gamma_1+\gamma_2) }  
 \end{pmatrix},
 \label{local u definition}
\end{equation}
for some rotation angles $\gamma_k$.
The local-$z$ equivalence class $\{U\}$ corresponding to $U$ is the set of elements $u_{\rm post} \, U \, u_{\rm pre}$ for all  $u_{\rm pre}, u_{\rm post}$. For a given gate $U$, $\{U\}$ typically occupies a four-dimensional manifold, depending on four rotation angles. But because (\ref{CZ definition}) is diagonal, $\{{\sf CZ}\}$ instead forms a two-dimensional sheet,
\begin{equation}
\{ {\sf CZ} \} = {\rm phase \ factor} \times 
\begin{pmatrix}
 1 & 0 & 0 & 0 \\
 0 & e^{-i \gamma_2} & 0 & 0 \\
 0 & 0 & e^{-i \gamma_1}  & 0 \\
 0 & 0 & 0 & -e^{-i (\gamma_1+\gamma_2) }  
\end{pmatrix}.
\label{CZ sheet}
\end{equation}
The {\sf CZ} gate (\ref{CZ definition}) can be obtained by reaching any point in the $\{{\sf CZ}\}$ plane and then performing auxiliary $z$ rotations. And it is straightforward to confirm that \cite{DiCarloNat09}
\begin{equation}
\begin{pmatrix}
 -1 & 0 & 0 & 0 \\
 0 & 1 & 0 & 0 \\
 0 & 0 & 1 & 0 \\
 0 & 0 & 0 & 1 
 \end{pmatrix}
\circeq
\begin{pmatrix}
 1 & 0 & 0 & 0 \\
 0 & -1 & 0 & 0 \\
 0 & 0 & 1 & 0 \\
 0 & 0 & 0 & 1 
 \end{pmatrix}
\circeq
\begin{pmatrix}
 1 & 0 & 0 & 0 \\
 0 & 1 & 0 & 0 \\
 0 & 0 & -1 & 0 \\
 0 & 0 & 0 & 1 
\end{pmatrix}
\circeq
\begin{pmatrix}
 1 & 0 & 0 & 0 \\
 0 & 1 & 0 & 0 \\
 0 & 0 & 1 & 0 \\
 0 & 0 & 0 & -1 
\end{pmatrix}.
\end{equation}
\end{widetext}
We note that bus rotations, which cannot be directly implemented with microwave pulses or frequency excursions, are compiled into future qubit rotations. 
 
The discussion above assumed a pair of qubits or a qubit and resonator, but it applies to a QVN processor in the interacting eigenfunction basis (\ref{QVN eigenstate}) after a minor modification. In the bare basis, the {\sf CZ} gate is typically defined through its action (\ref{CZ definition}) on a pair of qubits (or a qubit and resonator). Then, action on a bare computational basis state such as $ | q_1 q_2 \cdots q_n m_1 m_2 \cdots m_n b \rangle$ follows from the tensor-product form of that bare state. In the eigenstate basis the {\sf CZ} gate must be {\it defined} through its action on
\begin{equation}
\overline{| q_1 q_2 \cdots q_n m_1 m_2 \cdots m_n b \rangle},
\end{equation}
such as to reproduce the ideal action on the bare states to which they are perturbativey connected. For example, the {\sf CZ} gate on qubit $q_1$ and the bus acts ideally as
\begin{eqnarray}
{\sf CZ}  \, \overline{\big| 0 q_2  q_3 q_4 m_1 m_2 m_3 m_4 0 \big\rangle} &=&  \overline{\big| 0 q_2  q_3 q_4 m_1 m_2 m_3 m_4 0 \big\rangle} 
\nonumber \\
{\sf CZ}  \, \overline{\big| 0 q_2  q_3 q_4 m_1 m_2 m_3 m_4 1 \big\rangle} &=&  \overline{\big| 0 q_2  q_3 q_4 m_1 m_2 m_3 m_4 1 \big\rangle} 
\nonumber \\
{\sf CZ}  \, \overline{\big| 1 q_2  q_3 q_4 m_1 m_2 m_3 m_4 0 \big\rangle} &=&  \overline{\big| 1 q_2  q_3 q_4 m_1 m_2 m_3 m_4 0 \big\rangle} 
\nonumber \\
{\sf CZ}  \, \overline{\big| 1 q_2  q_3 q_4 m_1 m_2 m_3 m_4 1 \big\rangle} &=& - \, \overline{\big| 1 q_2  q_3 q_4 m_1 m_2 m_3 m_4 1 \big\rangle} ,
\nonumber 
\end{eqnarray}
where $q_i, m_i \in \lbrace 0, 1, 2, \dots \rbrace$.

\subsection{Fidelity definitions}
\label{fidelity definition section}

The gate or process fidelity measure we use in this work is based on a state fidelity defined by the inner product of the ideal and realized final (pure) states, squared. This leads to a state-averaged fidelity given by \cite{ZanardPRA04,PedersenPLA07}
\begin{equation} 
F_{\rm ave}\left(U,U_{\rm target}\right) \equiv \frac{{\rm Tr}(U^{\dagger} U)+
\big| {\rm Tr} \, (U_{\rm target}^{\dagger}U) 
\big|^{2}}{20},
\label{f}
\end{equation}
where $U$ is the realized time-evolution operator in the interacting eigenfunction basis after auxiliary $z$ rotations, projected into the relevant computational subspace, and $U_{\rm target}= {\sf CZ}$ [see (\ref{CZ definition})]. Note that the {\it projected} $U$ is not necessarily unitary here, and that the first term in (\ref{f}) characterizes the possible leakage from the computational basis (non-unitarity) whereas the second term is proportional to the square of the Hilbert-Schmidt inner product of $U$ with $U_{\rm target}$. Although $U$ is not assumed to be unitary, the expression (\ref{f}) assumes a pure state and is (obviously) not valid in the presence of decoherence. [The formula (\ref{f}) assumes that the Kraus representation for the completely positive process is not necessarily trace preserving, but it has only one term.] The form (\ref{f}) also assumes an average over a 4-dimensional Hilbert space;  in the $N$-dimensional generalization the denominator is $N+N^2$, which is necessary (note numerator) to assure that $F_{\rm ave}\left(U_{\rm target} ,U_{\rm target} \right)$ is unity.

It is also useful to calculate the minimum or worst-case fidelity. The minimum fidelity of interest here is the state fidelity {\it minimized} over initial computational states, for a gate that has already been optimized (by maximizing $F_{\rm ave}$). In Sec.~\ref{Strauch CZ gate section} we argued that the dominant intrinsic error mechanism (for an optimal pulse) in the truncated qubit-resonator model (\ref{qb hamiltonian}) is the nonadiabatic excitation of the $\overline{|02\rangle}$ channel, in which there are two photons left in the bus resonator. Therefore, in the model (\ref{qb hamiltonian}), the minimum state fidelity occurs for the initial eigenstate $\overline{|11\rangle}$. In the ${\rm QVN_4}$ processor, this worst-case state is written [in the notation of (\ref{QVN eigenstate})] as 
\begin{equation}
\overline{\ket{1 000 0000 1}},
\label{min fidelity eigenstate}
\end{equation}
where we have assumed a {\sf CZ}  gate between qubit $q_1$ and the bus.
Numerical simulation of this gate in the ${\rm QVN_4}$ processor confirms that the minimum {\sf CZ} fidelity indeed occurs for the initial state (\ref{min fidelity eigenstate}), and is due to leakage from the computational subspace. We therefore define the minimum fidelity to be the state fidelity for initial condition (\ref{min fidelity eigenstate}), 
\begin{equation} 
F_{|11\rangle} \equiv \big| \overline{\bra{1 000 0000 1}}  U 
\overline{\ket{1 000 0000 1}} \big|^2.
\label{F11 definition}
\end{equation}
Note that this expression is not sensitive to the value of the controlled phase, and only accounts for leakage from the computational subspace. Simulated values of $F_{|11\rangle}$ for the qubit-bus {\sf CZ} gate are given in Table \ref{fidelity table}, along with estimates of this same quantity that are discussed below in Sec.~\ref{fidelity estimator section}.

\subsection{Pulse shape}
\label{pulse shape section}

In the QVN Hamiltonian (\ref{QVN hamiltonian}), the qubit frequencies $\epsilon_i$ are the only available experimental controls. [There are also single-qubit terms for microwave pulses that are not shown in (\ref{QVN hamiltonian}) and not used in this work.]  During a {\sf CZ} gate between a given qubit and the resonator bus, the frequency of that qubit is varied according to
(\ref{pulseEq}), where
\BEq
{\rm Erf}(t) \equiv \frac{2}{\sqrt{\pi}}\int_{0}^{t}e^{-x^{2}}dx,
\EEq 
with the other qubits remaining at the parking frequency $\omega_{\rm park}$ (given in Table \ref{parameter table}). 

Two quantities related to the pulse switching---$\sigma$ and $t_{\rm ramp}$---appear in the pulse profile function (\ref{pulseEq}). We do this to emphasize that, in principal, two independent quantities could be used to quantify the shape of the switching profile. The first is the time duration (or width) of the switching, characterized by the standard deviation $\sigma$. The second is the trunction time of the pulse, measured from the center (half maximum) of the frequency switch, which is equal to $t_{\rm ramp}/2$. However in this work we always use the (somewhat arbitrary) relation (\ref{tramp specification}), which amounts to cutting off the pulse at $2\sqrt{2}$ standard deviations from the switching midpoints.

If the pulse shape function (\ref{pulseEq}) is used only when $0 \le t \le t_{\rm gate}$, and $\epsilon$ is set to $\omega_{\rm off}$ otherwise, there will be small pulse discontinuities at $t \! = \! 0$ and $t_{\rm gate}$, the size of which is determined by the condition (\ref{tramp specification}). Assuming $t_{\rm ramp} \ll t_{\rm gate},$ we have
\BEq\label{pulse truncation}
\epsilon(0) = \epsilon(t_{\rm gate}) \approx \omega_{\rm off} + \frac{\omega_{\rm on}-\omega_{\rm off}}{2}
\bigg[ 1 - {\rm Erf}\big(2\big) \bigg],
\EEq
which differs from the asymptotic value $\omega_{\rm off}$ by an amount
\begin{equation}
\frac{\omega_{\rm on}-\omega_{\rm off}}{2}
\bigg[ 1 - {\rm Erf}\big(2\big) \bigg],
\label{pulse discontinuity}
\end{equation}
where $1 - {\rm Erf}(2) \approx 0.5\%$. However, in an experiment these discontinuities are usually smoothed over by additional pulse shaping. Moreover, our simulations begin $t \! = \! 0$ and end at $t \! = \!  t_{\rm gate}$, so the truncation only slightly affects the pulse shape: The initial and final detuned qubit frequency is actually a few MHz smaller than $\omega_{\rm off}$.

Having fixed the relation (\ref{tramp specification}), there are then two measures of the pulse switching time, $\sigma$ and $t_{\rm ramp}$, with $t_{\rm ramp}$ providing a convenient measure of the time duration of the ramps. This property can be seen in the pulse shape example of Fig.~\ref{fig:pulse}.
 
\subsection{Gate optimization}
\label{optimization section}

We numerically optimize the two pulse-shape control parameters $\omega_{\rm on}$ and $t_{\rm on}$, as well as the two auxiliary $z$ rotation angles, $\gamma_1$ and $\gamma_2$, to maximize the average gate fidelity (\ref{f}). All other pulse parameters ($\omega_{\rm off}$, $\sigma$, $t_{\rm ramp}$) are fixed. The values of $\omega_{\rm off}$ and $t_{\rm ramp}$ are determined by the system optimization analysis of Sec.~\ref{system optimization section}, and then $\sigma$ is obtained through relation (\ref{tramp specification}). The roles played by the control parameters $\omega_{\rm on}$ and $t_{\rm on}$ are discussed above in Sec.~\ref{Strauch CZ gate section}.

The fidelity optimization procedure is carried out in two stages: In the first stage we take $\omega_{\rm on}$ to be equal to its approximate value [see (\ref{omega_on formula for simplified model})]
\begin{equation}
\omega_{\rm b} + \eta,
\label{perturbative won}
\end{equation}
and optimize $t_{\rm on}$ to get close to the two-dimensional equivalence class $\{ {\sf CZ} \}$ defined in (\ref{CZ sheet}). We do this by minimizing a sum of two positive errors, one measuring the deviation of the absolute values of the matrix elements of the evolution operator $U$ (projected into the $q_1$--$b$ subspace) from that of the four-dimensional identity matrix, the other measuring the deviation from the ideal relationship between the phases of the diagonal elements indicated in (\ref{CZ sheet}). This first stage yields an approximate value of $t_{\rm on}$, as well as approximate rotation angles
\begin{eqnarray}
\gamma_1 & \approx & \arg \, \overline{\langle 10|} U \overline{|10\rangle}
- \arg \, \overline{\langle 00|} U \overline{|00\rangle} ,\\
\gamma_2 & \approx & \arg \, \overline{\langle 01|} U \overline{|01\rangle}
- \arg \, \overline{\langle 00|} U \overline{|00\rangle} .
\label{approximate rotation angles formula}
\end{eqnarray}

In the second stage of optimization, we use the approximate values of $\omega_{\rm on}$, $t_{\rm on}$, $\gamma_1$, and $\gamma_2$, obtained from the first stage, as seeds for a full four-dimensional ($\omega_{\rm on}$, $t_{\rm on}$, $\gamma_1$, $\gamma_2$) nonlinear maximization of the average fidelity (\ref{f}) between 
\begin{equation}
u(\gamma_1,\gamma_2)\times U
\end{equation}
and the standard {\sf CZ} gate (\ref{CZ definition}). Here $U$ is the projected evolution operator and $u \in {\rm SU(2)}\otimes{\rm SU(2)}$ is defined in (\ref{local u definition}). 

There are two obvious approaches to simulating the gate evolution in the interacting eigenfunction basis (\ref{QVN eigenstate}). The first is to use the appropriate $H_{\rm idle}$ eigenstates to transform the target gate (\ref{CZ definition}) to the bare basis, and perform the actual simulation in the bare basis. However this approach requires that the subspace projection operators---which are naturally defined in the eigenstate basis---also be transformed. The second approach, which we follow here, is to transform the QVN Hamiltonian (\ref{QVN hamiltonian}) to the eigenstate basis and perform the simulation in that basis.

\subsection{Switching error and fidelity estimate}
\label{fidelity estimator section}

In this section we calculate the transition probability caused by a change of the qubit frequency during a {\sf CZ} pulse, {\sf MOVE} gate, or any other operation in the QVN system or related superconducting architectures. The problem will be treated quite generally and then applied to the Strauch {\sf CZ} gate of Table \ref{fidelity table}, resulting in a simple fidelity estimator for that gate. The quantities $A$, $p_{\rm sw}$, and $F_{|11\rangle}^{\rm (est)}$ appearing in Table \ref{fidelity table} are discussed in this section.

Imagine that we have prepared an initial interacting system eigenfunction $\overline{|a\rangle}$ prior to performing a {\sf CZ} operation or other gate that involves changing the frequency of one or more qubits. We assume that the ideal (target) behavior during the frequency switch or ramp itself is the {\it identity} map (times a phase factor), and that the $\overline{|a\rangle}$ channel does not cross any others in the system. The population loss during the ramp will therefore be exponentially suppressed if the switching time is long enough.

In a multi-qubit system there is typically a large number of nonresonant channels coupled to $\overline{|a\rangle}$ that can be excited by the frequency switch. However, when the ramp fidelity is high and the probabilities of the undesired transitions
\begin{equation}
\overline{|a\rangle} \rightarrow \overline{|b\rangle}, \overline{|b'\rangle}, \overline{|b''\rangle},
\cdots
\end{equation}
are small, they can be individually estimated perturbatively (neglecting 
interference), thereby reducing the problem to a sum of independent 
two-channel problems
\begin{equation}
\begin{matrix}
\overline{|a\rangle} && \longrightarrow &&  \overline{|b\rangle} \\
\overline{|a\rangle}  && \longrightarrow &&  \overline{|b'\rangle} \\
\overline{|a\rangle}  && \longrightarrow &&  \overline{|b''\rangle} \\
&& \vdots  && \\
\end{matrix}
\end{equation}
each characterized by a time-dependent detuning $\Delta$ and a coupling $G$. Without loss of generality we can shift the energy of a given two-channel problem so that the bare final state has zero energy. Each nonadiabatic transition can therefore be described by a general two-channel model of the form
\begin{equation}
H = \begin{pmatrix}
\Delta(t)  & G  \\
G & 0 \\
\end{pmatrix},
\label{2-channel model}
\end{equation}
in the bare basis spanned by $\lbrace |a\rangle, |b\rangle \rbrace$. The undesired final state $|b\rangle$ has a fixed energy 0 and the energy of $|a\rangle$ varies in time with detuning $\Delta$. The coupling $G$ is assumed to be a real, positive constant. 

The instantaneous eigenstates of (\ref{2-channel model}) are
\begin{eqnarray}
\overline{|a\rangle} &=& \cos \frac{\chi }{2}|a\rangle + \sin \frac{\chi }{2} |b\rangle , 
\label{instantaneous eigenstate i} \\
\overline{|b\rangle} &=& \cos \frac{\chi }{2} |b\rangle - \sin \frac{\chi }{2} |a\rangle ,
\label{instantaneous eigenstate f}
\end{eqnarray}
where
\begin{equation}
\chi \equiv \arctan \bigg( \frac{2G}{\Delta} \bigg).
\end{equation}
The instantaneous energies are
\begin{equation}
E_a= \frac{\Delta}{2}  + \sqrt{ \bigg( \frac{\Delta}{2} \bigg)^{\! 2} + G^2} 
\end{equation}
and
\begin{equation}
E_b  = \frac{\Delta}{2}  - \sqrt{ \bigg( \frac{\Delta}{2}  \bigg)^{\! 2}  + G^2} . 
\end{equation}
The $\overline{|b\rangle}$ channel is initially unoccupied at time $t\!=\!0$, and we are interested in the probability $p_{\rm sw}$ that the system, prepared in $\overline{|a\rangle}$, is found in $\overline{|b\rangle}$ after changing the detuning $\Delta$ from one value to another. We refer to this probability as the nonadiabatic switching error, which we calculate by expanding the wave function in the basis of instantaneous eigenstates 
(\ref{instantaneous eigenstate i}) and (\ref{instantaneous eigenstate f}), as
\begin{equation}
\big|\psi\big\rangle = \sum_{m=a,b} \psi_m \, e^{-i \int_0^t E_m \, d\tau} \, \overline{\big| m \big\rangle}.
\end{equation}
This leads to
\begin{equation}
\frac{d\psi_b}{dt} = -  e^{-i \int_0^t (E_a-E_b) \, d\tau} \, \overline{\big\langle b \big|} \frac{\partial}{\partial \Delta} \overline{\big| a \big\rangle} \  \frac{d\Delta}{dt}  \ \psi_a,
\label{exact final state equation}
\end{equation}
where 
\begin{equation}
\overline{\big\langle b \big|} \frac{\partial}{\partial \Delta} \overline{\big| a \big\rangle} = - \frac{G}{ \Delta^2} \times
\frac{1}{1+\big(\frac{2G}{\Delta}\big)^2}.
\label{exact nonadiabatic matrix element}
\end{equation}
The nonadiabatic matrix element (\ref{exact nonadiabatic matrix element}) has been written so that the second term approaches unity in the $G \ll \Delta$ perturbative limit.

At time $t \! = \! 0$, $\psi_a = 1$. An approximate expression for 
\begin{equation}
p_{\rm sw} \equiv |\psi_b(t_{\rm final})|^2
\end{equation}
can be obtained from (\ref{exact final state equation}) by assuming that $|\psi_b| \ll 1$ throughout the evolution, so that $\psi_a \approx 1$ for all $t$. Then
\begin{equation}
p_{\rm sw} \! = \!  \bigg| \! \int  \! \frac{G {\dot \Delta} \,
e^{- i \int_0^t \! \Omega \, d\tau} }{\Omega^2} dt \bigg|^2 \! \!
= \frac{1}{4} \, \bigg| \! \int  \! {\dot \chi} \, e^{- i \int_0^t \! \Omega \, d\tau} 
dt \bigg|^2 \! \! ,
\label{leakage integral}
\end{equation}
where
\begin{equation}
\Omega \equiv E_a-E_b = \sqrt{\Delta^2 + 4 G^2}
\end{equation}
is the instantaneous splitting. We can simplify (\ref{leakage integral}) further by assuming $G \ll \Delta$, which will be the case for the applications considered below. In this perturbative limit we therefore obtain
\begin{equation}
p_{\rm sw} = \bigg| \int  \frac{G {\dot \Delta}}{\Delta^2}  e^{-i \int_0^t \Delta d\tau} \, dt \ \bigg|^2 \! \! .
\label{perturbative leakage integral}
\end{equation}
We emphasize that the form (\ref{perturbative leakage integral}) assumes that $\Delta$ does not pass through zero, which would cause Landau-Zener tunneling and invalidate the perturbative analysis.

\begin{figure}[htb]
\centering
\includegraphics[angle=0,width=0.95\linewidth]{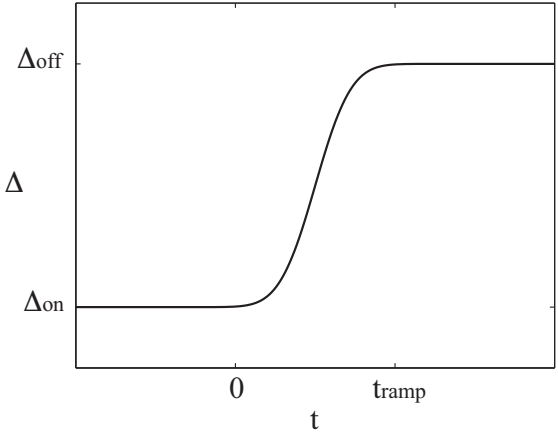}
\caption{Detuning pulse profile (\ref{single switch profile}) for a single frequency switch.}
\label{F figure}
\end{figure}

In this work we are specifically interested in $p_{\rm sw}$ for a single switch of the detuning from $\Delta_{\rm on}$ to $\Delta_{\rm off}$ (or the reverse) according to the smooth, error-function based profile
\begin{equation}
\Delta(t) = \frac{\Delta_{\rm off}+\Delta_{\rm on}}{2} + \frac{\Delta_{\rm off}-\Delta_{\rm on}}{2} \,
{\rm Erf}\left( \frac{t- \frac{1}{2} t_{\rm ramp}}{ \sqrt{2} \sigma  }\right) ,
\label{single switch profile}
\end{equation}
shown in Fig.~\ref{F figure}. The standard deviation $\sigma$ characterizes the switching time of the pulse and $t_{\rm ramp}$ [related to $\sigma$ through (\ref{tramp specification})] specifies its truncation, as discussed in Sec.~\ref{pulse shape section}. We use this switching profile for both ${\sf CZ}$ and ${\sf MOVE}$ gates. The switching error for the single switch profile defined in (\ref{single switch profile}) and shown in Fig.~\ref{F figure} can be expressed as
\begin{equation}
p_{\rm sw} = \bigg( \frac{G}{\Delta_{\rm on}} \bigg)^2  \, \big| A  \big|^2 \! ,
\label{switching error formula}
\end{equation}
where
\begin{equation}
A\big(\Delta_{\rm on},\Delta_{\rm off},\sigma \big) \equiv \Delta_{\rm on} \int_0^{t_{\rm ramp}} \! \!  \frac{{\dot \Delta}}{\Delta^2}  \, e^{-i \int_0^t \Delta \, d\tau} \, dt .
\label{A integral}
\end{equation}

\begin{widetext}

\begin{figure}[htb]
\centering
\includegraphics[angle=0,width=0.90\linewidth]{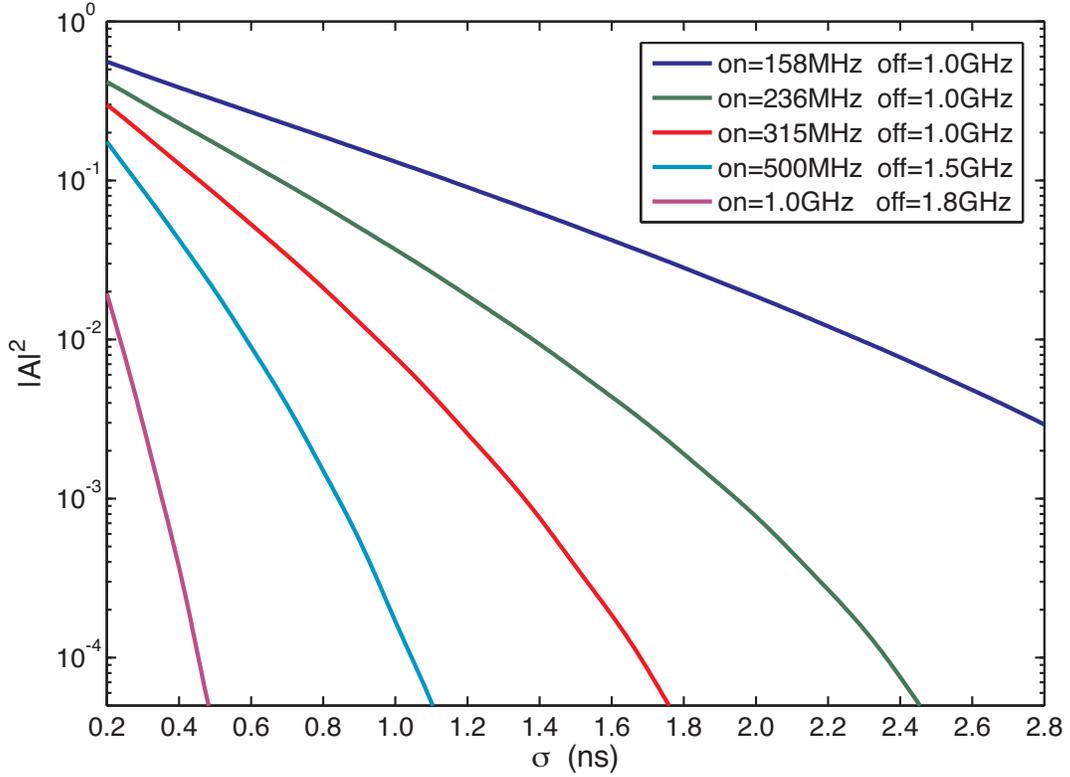}
\caption{(Color online) $|A|^2$ versus switching time $\sigma$ for indicated values of 
$\Delta_{\rm on}$ and $\Delta_{\rm off}$.}
\label{Asquared figure}
\end{figure}

\end{widetext}

The dimensionless quantity $|A|^2$ is plotted in Fig.~\ref{Asquared figure} for five instances of $\Delta_{\rm on}$ and $\Delta_{\rm off}$ relevant to this work. We note that $p_{\rm sw}$ evidently decreases as an exponential function of $\sigma$, as expected for a nonadiabatic process. However, the dependence of $|A|^2$ on $\sigma$ for large $\sigma$ is somewhat intricate, a consequence of the error-function ramp shape. For very large values of $\sigma$---not shown in Fig.~\ref{Asquared figure}---the decay of $|A|^2$ becomes slower (the location of the crossover depends on the details of the pulse truncation). Although an approximate analytic expression for $|A|^2$ can be derived for this large-$\sigma$ limit, the formula is not useful for the regimes of interest here.

We now turn to the application of the switching error formula (\ref{switching error formula}) to the {\sf CZ} gate of Table \ref{fidelity table}. As we have explained above in Sec.~\ref{fidelity definition section}, the initial condition in the ${\rm QVN}_4$ processor with the minimum fidelity after a {\sf CZ} gate (between qubit $q_1$ and the bus) is the eigenstate (\ref{min fidelity eigenstate}). For a perfect {\sf CZ} gate this state would map to
\begin{equation}
-\overline{|1 000 0000 1 \rangle},
\end{equation}
but in practice we will find population in other channels as well, the dominant error (for an optimal pulse) occuring in $\overline{|0 000 0000 2 \rangle}$, which has two excitations in the bus. This dominant error mechanism involves only a single qubit and resonator, and can be understood entirely within the truncated model (\ref{qb hamiltonian}). In the language of that model, where the bare states (interacting eigenstates) are written as $|qb\rangle$ ($\overline{|qb\rangle}$), the dominant fidelity loss of the $\overline{|11\rangle}$ eigenstate is caused by nonadiabatic leakage to $\overline{|02\rangle}$. We call this leakage error $\mathbb{E}_2$.

It is possible to understand this dominant $\overline{\ket{11}} \rightarrow \overline{\ket{02}}$ leakage error in a further simplified two-channel model that only includes the channels $\overline{|11\rangle}$ and $\overline{|02\rangle}$. Given the strong interaction of the bare $|11\rangle$ state with $|20\rangle$, it is not at all obvious that such a simplification is possible. However, during most of the switching, the detuning between $|11\rangle$ and $|20\rangle$ is much larger than their interaction strength $\sqrt{2} g_{\rm b}$, so they are effectively decoupled. (And while the qubit sits at the $\omega_{\rm on}$ frequency, the Hamiltonian is time-independent.) Therefore a two-channel description should be possible, although it will slightly overestimate the excitation of $\overline{\ket{02}}$. Numerical investigation confirms that the $\overline{|20\rangle}$ channel can indeed be disregarded except for the level repulsion it produces on the $\overline{|11\rangle}$ state (see below), which is crucial for obtaining an accurate fidelity estimate.

In the bare $\lbrace |11\rangle, |02\rangle \rbrace$ basis, (\ref{qb hamiltonian}) reduces to
\begin{equation}
H = \begin{pmatrix}
\epsilon & \sqrt{2}g_{\rm b}   \\
\sqrt{2}g_{\rm b}  & \omega_{\rm b}  \\
 \end{pmatrix} + {\rm const.}
\end{equation}
At the beginning of the {\sf CZ} pulse of Table \ref{fidelity table}, the qubit frequency is  $1.0 \, {\rm GHz}$ above the bus. $\epsilon$ then decreases to $\omega_{\rm on} \approx \omega_{\rm b} + \eta$ and returns to the detuned configuration in the manner of Fig.~\ref{fig:pulse}. The total leakage to $\overline{\ket{02}}$ can be estimated as twice---because there are two switching events, which we assume to contribute incoherently---the value of $p_{\rm sw}$. Therefore the error $\mathbb{E}_2$ introduced in (\ref{CZ evolution 11 component with switching error}) is given by
\begin{equation}
\mathbb{E}_2 = 2 p_{\rm sw}.
\end{equation} 
To evaluate the switching error in this case, we use (\ref{switching error formula}) with the parameter values
\begin{eqnarray}
G &=& \sqrt{2} g_{\rm b},  \label{G value} \\ 
\Delta_{\rm on} &=& \eta - \sqrt{2} g_{\rm b}, \label{Delta on value} \\
\Delta_{\rm off} &=& \omega_{\rm off}  - \omega_{\rm b}. 
\end{eqnarray}
The value of $\Delta_{\rm on}$ accounts for the level repulsion caused by the neglected $|20\rangle$ state, which causes the $\overline{|11\rangle}$ channel to shift downwards towards $\overline{|02\rangle}$; this large effect is evident in Fig.~\ref{fig:energylevels}. [We ignore here a smaller repulsion by $\overline{|02\rangle}$, which would lead to the addition of a small positive correction to (\ref{Delta on value}).] $\Delta_{\rm off}/2\pi$ is always $1.0 \, {\rm GHz}$ for the gates listed in Table \ref{fidelity table}. The required $|A|^2$ values are obtained from Fig.~\ref{Asquared figure} and are provided in Table \ref{fidelity table} along with $p_{\rm sw}$.

The minimum fidelity estimate
\begin{equation}
F_{|11\rangle}^{(\rm est)} \equiv 1-\mathbb{E}_2 =  1 -2 \, p_{\rm sw}
\label{finimum fidelity estimate}
\end{equation}
is also given in Table \ref{fidelity table} and compared to simulated ${\rm QVN}_4$ values of $F_{\rm ave}$ and $F_{|11\rangle}$. We find that (\ref{finimum fidelity estimate}) is a reliable predictor of the worst-case fidelity $F_{|11\rangle}$ in ${\rm QVN}_4$, confirming that the nonadiabatic switching error is the dominant fldelity loss mechanism here. Although this error will always be present,  it can be exponentially suppressed by increasing the switching time. 
 
Finally, we briefly comment on the nonadiabatic switching errors between the $\overline{|01\rangle}$ and $\overline{|10\rangle}$ eigenstates, which we have argued to be subdominant to the excitation of $\overline{|02\rangle}$, but which naively are of the same order. There are two reasons why the $\overline{|01\rangle}$ and $\overline{|10\rangle}$ switching errors are considerably smaller: First, the matrix element coupling $\overline{|01\rangle}$ and $\overline{|10\rangle}$ is a factor of $\sqrt{2}$ smaller than that between $\overline{|11\rangle}$ and $\overline{|02\rangle}$, and this factor gets squared in (\ref{switching error formula}). And the second---but quantativitely more important---reason is that while level repulsion considerably enhances the $\overline{|02\rangle}$ excitation [recall (\ref{Delta on value})], it (slightly) suppresses transitions between $\overline{|01\rangle}$ and $\overline{|10\rangle}$. 

We can estimate the switching errors between the $\overline{|01\rangle}$ and $\overline{|10\rangle}$ channels during a {\sf CZ} gate by using (\ref{switching error formula}) with parameters
\begin{eqnarray}
G &=& g_{\rm b},  \label{G value lower band} \\ 
\Delta_{\rm on} &=& \eta + \frac{2 g_{\rm b}^2}{\eta}, \label{Delta on value lower band}
\end{eqnarray}
and with $\Delta_{\rm off}/2 \pi = 1 \, {\rm GHz}$ as before. The expression (\ref{Delta on value lower band}) for $\Delta_{\rm on}$ accounts for the level repulsion between $\overline{|01\rangle}$ and $\overline{|10\rangle}$, which suppresses the switching error, in contrast with the strong enhancement indicated in (\ref{Delta on value}). Considering, for example, the 99.9\% {\sf CZ} gate of Table \ref{fidelity table} designed for the $\eta/2\pi = 300 \, {\rm MHz}$ qubit, we find that $|A|^2 = 2.2 \! \times \! 10^{-3}$ and 
\begin{equation}
p_{\rm sw} = 4.4 \! \times \! 10^{-5}.
\label{lower band switching error}
\end{equation}
The total error in this case is $\mathbb{E}_1 = 2 p_{\rm sw}$. The estimate (\ref{lower band switching error}) for the switching error between $\overline{|01\rangle}$ and $\overline{|10\rangle}$ is more than an order-of-magnitude smaller than that between $\overline{|11\rangle}$ and $\overline{|02\rangle}$, which is $p_{\rm sw} = 1.2 \! \times \! 10^{-3}$ (see Table \ref{fidelity table}).

\subsection{Pulse shape errors}
\label{pulse shape errors section}

In Sec.~\ref{fidelity estimator section} we discussed the intrinsic error of the qubit-bus {\sf CZ} gate---assuming an optimal pulse shape---and identified its dominant source as a nonadiabatic switching error $\mathbb{E}_2$. In this section we discuss and quantify the fidelity loss caused by pulse shape and auxiliary $z$ rotation errors. By a pulse shape error we mean that the correct functional form (\ref{pulseEq}) is seen by the qubit, but with values of the parameters $t_{\rm on}$ and $\omega_{\rm on}$ that deviate from the optimal values. It is possible to develop simple analytic models for these error mechanisms (supported by numerical simulation) by noting that when the fidelity is very close to unity, the different error mechanisms present contribute independently and can be calculated separately. We then use these results to estimate the experimental pulse-control precision required during the implementation of a given $99.9\%$ or $99.99\%$ {\sf CZ} gate to keep any accompanying pulse shape error less than the base $10^{-3}$ or $10^{-4}$ gate error.

The simplest situation to consider is that where the correct values of $t_{\rm on}$ and $\omega_{\rm on}$ are used, but where the local $z$ rotation angles $\gamma_k \ (k=1,2)$ applied experimentally deviate from their optimal values by amounts $\varphi_k \ll 1.$  We estimate the resulting fidelity loss by imagining that we have achieved a perfect {\sf CZ}-class gate 
\begin{equation}
U= 
\begin{pmatrix}
 1 & 0 & 0 & 0 \\
 0 & e^{i \Theta_2} & 0 & 0 \\
 0 & 0 & e^{i \Theta_1}  & 0 \\
 0 & 0 & 0 & -e^{i (\Theta_1+ \Theta_2) }  
\end{pmatrix},
\label{CZ class gate}
\end{equation}
for some phase angles $\Theta_k$, but then apply $z$ rotation angles
\begin{equation}
\gamma_k = \Theta_k + \varphi_k
\end{equation}
that have errors $\varphi_k$. From (\ref{f}) we find that this leads to a leading order error $\mathbb{E} \equiv 1 - F_{\rm ave}$ given by
\begin{equation}
\mathbb{E} = \frac{ \varphi_1^2 + \varphi_2^2}{5}.
\label{fidelity loss from z rotation angle errors}
\end{equation}

Next we consider $t_{\rm on}$ and $\omega_{\rm on}$ errors. An error in either $t_{\rm on}$ or $\omega_{\rm on}$ has two consequences, the first is to modify the accumulated phases $\Theta_k$ in (\ref{CZ class gate}), and the second is to cause population and phase errors on the $\overline{|11\rangle}$ channel. Therefore we consider two types of pulse shape errors, the first where $t_{\rm on}$ or $\omega_{\rm on}$ is changed with no compensating changes in the auxiliary $z$ rotation angles, and the second where the $\gamma_k$ are reoptimized. 

In the first case, in which the error is clearly the largest, the resulting error is dominated by the $z$ rotation angle error itself, which can be estimated from (\ref{fidelity loss from z rotation angle errors}). Changing $t_{\rm on}$ by an amount $\delta t_{\rm on}$, or $\omega_{\rm on}$ by an amount $\delta\omega_{\rm on}$, changes the accumulated phase of the qubit (recall discussion of the qubit reference frame in Secs.~\ref{sec:model} and \ref{eigenstate basis section}) by
\begin{equation}
\varphi_1 = (\omega_{\rm on}-\omega_{\rm off}) \, \delta t_{\rm on} + \delta\omega_{\rm on} \, t_{\rm on} ,
\label{delta gamma1 uncompensated error}
\end{equation}
and that of the bus by
\begin{equation}
\varphi_2 = 0.
\label{delta gamma2 uncompensated error}
\end{equation}
We note that $\delta t_{\rm on}$ and $\delta\omega_{\rm on}$ can be positive or negative here, and that the total gate time $t_{\rm gate}$ is also (slightly) changed by $\delta t_{\rm on}$. The additional accumulated phase (\ref{delta gamma1 uncompensated error}) can be regarded as a rotation angle error because, by assumption, it is not compensated by the applied $z$ rotations (hence the notation). The error angle $\varphi_2$ is zero because of our choice of the local clock/reference frame for a resonator. Therefore, an error in either $t_{\rm on}$ or $\omega_{\rm on}$ with no compensating adjustment of the auxiliary $z$ rotation angles leads to a leading-order fidelity loss of
\begin{equation}
\mathbb{E} = \frac{ (\omega_{\rm off}-\omega_{\rm on} )^2}{5}  \, \delta t_{\rm on}^2 
+ \frac{ t_{\rm on}^2 }{5} \, \delta\omega_{\rm on}^2.
\label{fidelity loss with uncompensated z rotations}
\end{equation}
For an order-of-magnitude estimate it is sufficient to approximate $\omega_{\rm on}$ here by $\omega_{\rm b}$ and $t_{\rm on}$ by $t_{\rm on}^{\rm sudden}$ [see (\ref{ton sudden definition})], leading to the simpler estimate
\begin{equation}
\mathbb{E}^\prime = \frac{ (\omega_{\rm off} - \omega_{\rm b}  )^2}{5}  \, \delta t_{\rm on}^2 
+ \frac{ (t_{\rm on}^{\rm sudden})^2 }{5} \, \delta\omega_{\rm on}^2.
\label{simplified fidelity loss with uncompensated z rotations}
\end{equation}
Considering the $99.9\%$ {\sf CZ} gate of Table \ref{fidelity table} designed for the $\eta/2\pi \! = \! 300 \, {\rm MHz}$ transmon, we estimate from (\ref{simplified fidelity loss with uncompensated z rotations}) that a $10 \, {\rm ps}$ error in $t_{\rm on}$ (or a $1 \, {\rm MHz}$ error in $\omega_{\rm on}/2\pi$) would lead to a fidelity loss of $7.9\! \times 10^{-4}$ (or $4.9 \! \times 10^{-4}$), whereas numerical simulation of the same error in ${\rm QVN_4}$, which includes all subdominant processes, yields $4.5 \! \times 10^{-4}$ (or $1.2 \! \times \! 10^{-3}$). 

Next we consider the case where there is an error in $t_{\rm on}$ or $\omega_{\rm on}$, but the auxiliary $z$ rotation angles are optimal. Here the analysis closely follows that of Sec.~\ref{Strauch CZ gate section}, which is based on the truncated qubit-resonator model (\ref{qb hamiltonian}). In this situation the fidelity loss is dominated by deviations from the ideal evolution
\begin{equation}
{\sf CZ} \, \overline{|11\rangle} = - \overline{|11\rangle} 
\label{ideal CZ action on 11 state}
\end{equation}
of the $\overline{|11\rangle}$ channel. Pulse shape errors will lead to both population and phase errors on the right-hand-side of (\ref{ideal CZ action on 11 state}). We therefore parameterize the nonideal {\sf CZ} gate by
\begin{equation}
U= {\rm phase \, factor} \times
\begin{pmatrix}
 1 & 0 & 0 & 0 \\
 0 & 1 & 0 & 0 \\
 0 & 0 & 1  & 0 \\
 0 & 0 & 0 &  - \! e^{i \delta} \cos \frac{\theta}{2}  \\
\end{pmatrix}.
\label{U after 11 error}
\end{equation}
In (\ref{U after 11 error}) we have assumed perfect auxiliary $z$ rotations and have neglected the subdominant errors in the $\overline{|00\rangle}$, $\overline{|01\rangle}$, and $\overline{|10\rangle}$ channels.  The population error has been written in terms of a rotation angle error $\theta$, which will be interpreted (see below) as the deviation from $2\pi$ of the rotation angle in the two-dimensional subspace spanned by $|11\rangle$ and $|20\rangle$. The expression (\ref{U after 11 error}) does not include switching errors because  we are evaluating the effect of pulse-shape errors on an otherwise perfect, large-$\sigma$ {\sf CZ} gate. The average fidelity loss $\mathbb{E} \equiv 1 - F_{\rm ave}$ associated with (\ref{U after 11 error}) is, to leading order,
\begin{equation}
\mathbb{E} = \frac{3}{20} \, \delta^2 + \frac{1}{16} \, \theta^2.
\label{fidelity loss from 11 state errors}
\end{equation} 

What remains is to express the controlled phase error $\delta$ and rotation error angle $\theta$ in terms of $\delta t_{\rm on}$ and $\delta\omega_{\rm on}$. This involves only the two channels $\overline{|11\rangle}$ and $\overline{|20\rangle}$ of the qubit-resonator model (\ref{qb hamiltonian}), and we will use the same small-$\sigma$ approximation used in Sec.~\ref{Strauch CZ gate section} for our analysis of the $\overline{|11\rangle}$ channel dynamics. In the $\lbrace |11\rangle , |20\rangle \rbrace$ basis, (\ref{qb hamiltonian}) can be written as
\begin{equation}
H =
\begin{pmatrix}
\epsilon + \omega_{\rm b} & \sqrt{2} g_{\rm b} \\
\sqrt{2} g_{\rm b} & 2\epsilon - \eta \\
\end{pmatrix}.
\label{general H for 2d subspace}
\end{equation}
The eigenstates of (\ref{general H for 2d subspace}) are
\begin{eqnarray}
\overline{|11\rangle} &=& \cos \frac{\zeta}{2}\, |11\rangle - \sin \frac{\zeta}{2}\, |20\rangle, \\
\overline{|20\rangle} &=& \cos \frac{\zeta}{2}\, |20\rangle + \sin \frac{\zeta}{2}\, |11\rangle,
\end{eqnarray}
where
\begin{equation}
\zeta \equiv \arctan \bigg( \frac{2 \sqrt{2} g_{\rm b}}{ \epsilon-\omega_{\rm on}} \bigg),
\end{equation}
and the energies are
\begin{eqnarray}
E_{11} &=& \epsilon + \omega_{\rm b}  + \frac{\epsilon-\omega_{\rm on}}{2} \nonumber \\
 &-& \sqrt{ \bigg(\frac{\epsilon-\omega_{\rm on}}{2}\bigg)^2 + \big( \sqrt{2} g_{\rm b} \big)^2 } , \\
E_{20}  &=& \epsilon + \omega_{\rm b} + \frac{\epsilon-\omega_{\rm on}}{2}
\nonumber \\ 
&+& \sqrt{ \bigg(\frac{\epsilon-\omega_{\rm on}}{2}\bigg)^2 + \big( \sqrt{2} g_{\rm b} \big)^2 }.
\end{eqnarray}
Here we have used the expression (\ref{omega_on formula for simplified model}) for $\omega_{\rm on}$, which is appropriate for the model (\ref{general H for 2d subspace}).

The analysis below follows that of the $\overline{|11\rangle}$ channel evolution given Sec.~\ref{Strauch CZ gate section}, except here we introduce a timing error $\delta t_{\rm on}$ and a tuning error $\delta\omega_{\rm on}$. Starting in the strongly detuned configuration with $\epsilon-\omega_{\rm on} \gg {\rm g}_b$ in the eigenstate $\overline{|11\rangle} \approx |11\rangle$, and quickly switching to $\epsilon = \omega_{\rm on} + \delta \omega_{\rm on}$, leaves the system in the state
\begin{equation}
 |11\rangle =  \cos \frac{\zeta_{\rm on}}{2}  \, \overline{|11\rangle} + \sin \frac{\zeta_{\rm on}}{2} \, \overline{|20\rangle}.
 \label{state after tuning}
\end{equation}
Here $\zeta_{\rm on} \equiv \arctan (2 \sqrt{2} g_{\rm b} / \delta\omega_{\rm on})$, and the eigenstates in (\ref{state after tuning}) are for $\epsilon = \omega_{\rm on} + \delta \omega_{\rm on}$. Note that (\ref{state after tuning}) reduces to (\ref{tuned nonstationary state}) in the $\delta\omega_{\rm on} \rightarrow 0$ limit. Evolution with $\epsilon$ fixed at $\omega_{\rm on} + \delta \omega_{\rm on}$ for a time 
\begin{equation}
\frac{2 \pi}{E_{20} - E_{11}} =
\frac{ \pi}{\sqrt{ \big( \delta \omega_{\rm on} / 2 \big)^2 + \big( \sqrt{2} g_{\rm b} \big)^2 }  }
\label{ideal ton}
\end{equation}
would implement a $2 \pi$ rotation in the $\lbrace |11\rangle , |20\rangle \rbrace$ subspace, returning to $|11\rangle$ with a phase shift that depends on $\delta \omega_{\rm on}$.
We intentionally introduce a $t_{\rm on}$ pulse shape error and instead evolve for a time
\begin{equation}
t =
\frac{ \pi}{\sqrt{ \big( \delta \omega_{\rm on} / 2 \big)^2 + \big( \sqrt{2} g_{\rm b} \big)^2 }  } + \delta t_{\rm on},
\label{ton error definition}
\end{equation}
after which we detune and find the final state
\begin{eqnarray}
&-& e^{i \delta} \cos \frac{\theta}{2} \, \big|11\big\rangle + 
{\rm phase \ factor} \times \sqrt{\mathbb{E}_\theta} \, \big|20\big\rangle\nonumber \\
\approx &-& e^{i \delta} \cos \frac{\theta}{2} \, \overline{\big|11\big\rangle} + 
{\rm phase \ factor} \times \sqrt{\mathbb{E}_\theta} \, \overline{\big|20\big\rangle}, \ \ \
\end{eqnarray}
where 
\begin{equation}
\delta = - \frac{\pi \, \delta\omega_{\rm on}}{2\sqrt{2} g_{\rm b}} 
\label{delta formula}
\end{equation}
and
\begin{equation}
\mathbb{E}_\theta \equiv \sin^2 \frac{\theta}{2}= 2 g_{\rm b}^2 \, \delta t_{\rm on}^2.
\label{p_theta formula}
\end{equation}
These expressions are valid to leading order in $\delta\omega_{\rm on}$ or $\delta t_{\rm on}$, neglecting cross terms. Here $\mathbb{E}_\theta$ is the probability of leakage to $\overline{|20\rangle}$ resulting from a $t_{\rm on}$ error, which, as discussed above, causes a rotation error of angle $\theta$. An alternative estimate for $\theta$ (and hence $\mathbb{E}_\theta$) is
\begin{equation}
\theta \approx \frac{\delta t_{\rm on} } {t_{\rm on}} \times 2 \pi,
\label{ton approimate error angle formula}
\end{equation}
which also gives (\ref{p_theta formula}) [after using the sudden limit result  (\ref{ton sudden definition}) for $t_{\rm on}$]. Note that the leakage error $\mathbb{E}_\theta$ is independent of $\delta\omega_{\rm on}$ (to this order), enabling the phase $\delta$ to be intentionally adjusted by varying $\omega_{\rm on}$ only. Doing this generates (approximately) gates of the form 
\begin{equation}
\begin{pmatrix}
 1 & 0 & 0 & 0 \\
 0 & 1 & 0 & 0 \\
 0 & 0 & 1  & 0 \\
 0 & 0 & 0 &  \! e^{i(\pi + \delta)} \\
\end{pmatrix} \! ,
\label{controlled-phase gate}
\end{equation}
for small $\delta$, with alternative controlled phases \cite{YamamotoPRB10}. Gates (approximately) of the form (\ref{controlled-phase gate}) with arbitrary---not necessarily small---values of $\delta$ can be implemented by varying both $\omega_{\rm on}$ and $t_{\rm on}$ away from their optimal values.

Referring again to the $99.9\%$ {\sf CZ} gate of Table \ref{fidelity table} designed for the $\eta/2\pi \! = \! 300 \, {\rm MHz}$ transmon, we estimate from (\ref{fidelity loss from 11 state errors}), (\ref{delta formula}), and (\ref{p_theta formula}) that a $10 \, {\rm ps}$ error in $t_{\rm on}$ (or a $1 \, {\rm MHz}$ error in $\omega_{\rm on}/2\pi$) with optimal $z$ rotations would lead to a fidelity loss of about $4.0\! \times 10^{-6}$ (or $9.1 \! \times 10^{-5}$), whereas numerical simulation of the same error in ${\rm QVN_4}$, which includes all subdominant processes, yields $3.0 \! \times 10^{-6}$ (or $1.1 \! \times \! 10^{-4}$).

Finally, it is interesting to use the above estimates to bound the magnitude of the allowable $t_{\rm on}$ and $\omega_{\rm on}$ errors, such that the resulting pulse shape errors (\ref{fidelity loss from 11 state errors}) are subdominant to the $10^{-3}$ or $10^{-4}$ base gate error. This is done in Table~\ref{pulse shape precision table}. For example, a 99.9\% {\sf CZ} gate from Table~\ref{fidelity table} with a $t_{\rm on}$ error of $160 \, {\rm ps}$ will have an additional intrinsic error of $10^{-3}$ (and a total error of about $2 \times 10 ^{-3})$. Current experimental limitations on the control of $t_{\rm on}$ and $\omega_{\rm on}$ are considerably better than that required to suppress pulse shape errors below the $10^{-4}$ level.

\begin{table}[htb]
\centering
\caption{{\sf CZ} pulse shape precision requirements. The bounds listed in 
the $ t_{\rm on}$ column assume that this is the only type of pulse parameter inaccuracy present, with an (estimated) error given in the first column, and 
that the auxiliary $z$ rotation angles are reoptimized and implemented perfectly. 
The $ \omega_{\rm on}$ bounds are defined analogously. The error $\mathbb{E}$ is defined in (\ref{fidelity loss from 11 state errors}).}
\begin{tabular}{|c|c|c|}
\hline
$\mathbb{E}$ & $t_{\rm on}$ precision   &  $\omega_{\rm on}$ precision  \\
\hline 
$10^{-3}$ & $160 \, {\rm ps}$   & $4 \, {\rm MHz}$ \\
$10^{-4}$ & $50 \, {\rm ps}$   & $1 \, {\rm MHz}$ \\
\hline 
\end{tabular}
\label{pulse shape precision table}
\end{table}

\subsection{{\sf CZ} gate with $99.999\%$ fidelity}
\label{5 nines section}

Higher fidelities are also possible with the pulse shape (\ref{pulseEq}). An example is provided in Table \ref{5 nines table} for the $300 \, {\rm MHz}$ qubit. For this design we did not perform a separate $g_{\rm b}$ optimization for this higher fidelity, but instead used the value from Table \ref{fidelity table} optimized for the lower fidelities.

\begin{table}[htb]
\centering
\caption{\label{5 nines table} Optimal ${\rm QVN}_4$ gate fidelity for a Strauch {\sf CZ} gate between qubit $q_1$ and the bus.}
\begin{tabular}{|c|c|c|c|c|c|c|c|}
\hline  
$\eta/2\pi$ & $g_{\rm b}/2\pi$ & $g_{\rm m}/2 \pi$ &$t_{\rm ramp}$  & $\sigma$ & $t_{\rm gate}$ & $F_{\rm ave}$   \\
\hline  
$300 \, {\rm MHz}$ &  $45 \, {\rm MHz}$   &  $100 \, {\rm MHz}$ & $13 \, {\rm ns}$ & $2.30 \, {\rm ns}$ &  $25.7 \, {\rm ns}$ & 99.999\%  \\ 
\hline
\end{tabular}
\end{table}

\section{ADDITIONAL {\sf CZ} GATES}
\label{other CZ gates section}

\subsection{{\sf CZ} between bus and qubit $q_4$}
\label{qubit 4 CZ gate section}

The main focus of this paper is the {\sf CZ} gate between qubit $q_1$ and the bus in the ${\rm QVN}_4$ processor. Results for the other qubits are very similar, with the worst case being $q_4$, because the detuning to memory during the gate is slightly smaller. We find that the intrinsic fidelity of the 99.9\% {\sf CZ} gate for the $300 \, {\rm MHz}$ qubit given in Table \ref{fidelity table} changes from 99.928\% to 99.925\% if qubit $q_4$ is used instead of $q_1$.

\subsection{{\sf CZ} between two qubits in ${\rm QVN}_4$}
\label{CZ between qubits section}

\begin{figure}[htb]
\centering
\includegraphics[angle=0,width=1.00\linewidth]{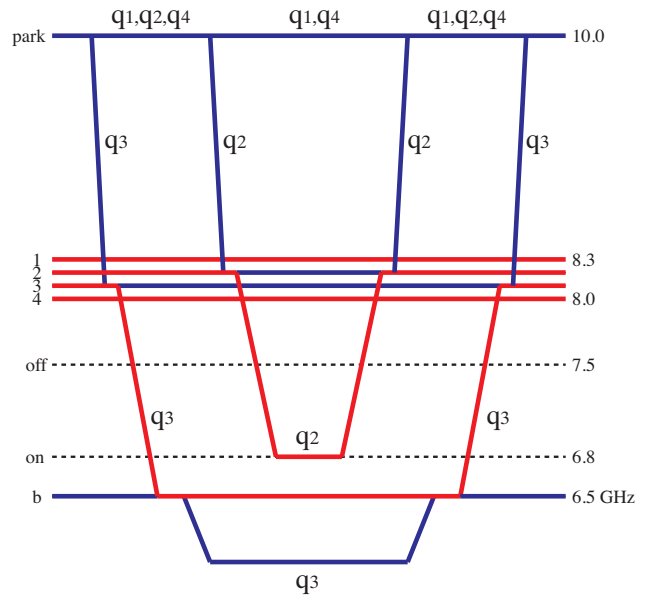}
\caption{(Color online) ${\rm QVN}_4$ mode diagram for the ${\sf CZ}_{23}$ gate. Gaussian filtering of the pulse is not shown.}
\label{mode diagram figure CZ23}
\end{figure}

In this section we explain how to perform a {\sf CZ} gate between two qubits---or more precisely, between two {\it memories}---in the QVN architecture. Such an operation is not elementary, as it can be composed of the qubit-bus {\sf CZ} combined with {\sf MOVE} gates. (There are also proposals for the direct implementation of a qubit-qubit {\sf CZ} gate in a QVN device \cite{GaliautdinovArxiv1103.4641,EggerPre12}.) Figure \ref{mode diagram figure CZ23} shows the experimental protocol for implementing the gate ${\sf CZ}_{23}$ between qubits $q_2$ and $q_3$, suppressing auxiliary $z$ rotations, and with all data starting and ending in memory. This mode diagram shows the time-dependence of all 9 device frequencies. The color indicates whether the qubit or resonator would be in the ground state (blue) or possibly excited state (red) in the weakly coupled limit. Modes are colored red if there is a finite occupation of the $|1\rangle$ state (in the weakly coupled limit), for some choice of initial conditions. All qubits are initially parked at the strongly detuned frequency of $10 \, {\rm GHz}$. Horizontal lines 1-4 represent memories, and $b$ is the bus. No red or red and blue lines with first-order {\it or} second-order couplings cross (to avoid Landau-Zener transitions), and no more than one qubit is occupied at any time (to avoid second-order qubit-qubit interactions mediated by the bus).

\begin{figure}[htb]
\centering
\includegraphics[angle=0,width=1.00\linewidth]{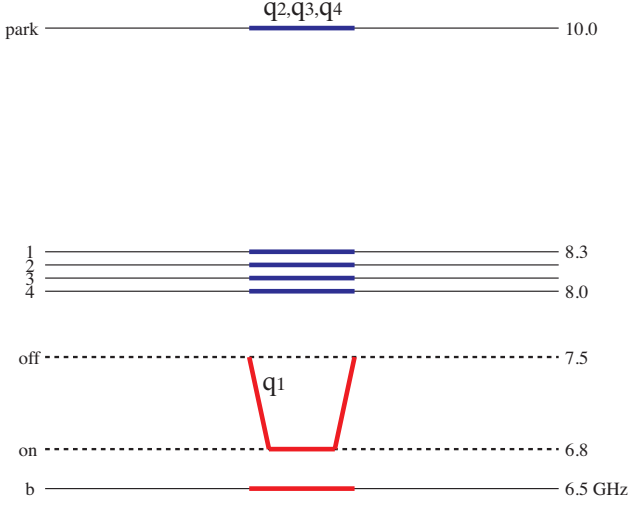}
\caption{(Color online) {\sf CZ} gate between $q_1$ and bus. This diagram describes the gate of Table \ref{fidelity table}.}
\label{mode diagram figure CZqb}
\end{figure}

Beginning with the four memory registers in an arbitrary (possibly entangled) state, the bus is loaded by a $5 \, {\rm ns}$ {\sf MOVE} gate from $m_3 \rightarrow q_3$ followed by a $10 \, {\rm ns}$ {\sf MOVE} to the bus. These are approximate gate times (time estimates for these gates are given in Sec.~\ref{system optimization section} and a concrete example is provided below). Qubit $q_2$ is then loaded and tuned to the frequency $\omega_{\rm on}$  determined by optimization. This central portion of the gate is close (but not exactly the same as) the qubit-bus {\sf CZ} gate of Table \ref{fidelity table}, which, for the purposes of comparison, is shown in Fig.~\ref{mode diagram figure CZqb}.

We simulated the gate ${\rm CZ}_{23}$ shown in Fig.~\ref{mode diagram figure CZ23}, using the 99.99\% {\sf CZ} of Table \ref{fidelity table} for a $300 \, {\rm MHz}$ qubit, and starting with the memory register in the GHZ state
\begin{equation}
\frac{ \overline{|0000\rangle} + \overline{|1111\rangle}}{\sqrt{2}}.
\label{cat state}
\end{equation}
The {\sf MOVE} gates also have fidelities around 99.99\%. Note that $H_{\rm idle}$ and the associated computational basis states (interacting eigenfunctions of $H_{\rm idle}$) are different at the beginning and end of each {\sf MOVE} gate. The total ${\sf CZ}_{23}$ gate time is 
\begin{equation}
t_{\rm gate} = 55 \, {\rm ns},
\label{CZ23 fidelity}
\end{equation}
and the final state fidelity (overlap squared) is
\begin{equation}
F= 99.94\%.
\label{CZ23 fidelity}
\end{equation}
In addition to the $23 \, {\rm ns}$ qubit-bus {\sf CZ} gate, there are four {\sf MOVE} to/from memory operations, each taking about $3.5 \, {\rm ns}$, and two {\sf MOVE} to/from bus gates, each taking about $9 \, {\rm ns}$. There are also local $z$ rotations (not shown) between each gate.

A few remarks about the encouraging result (\ref{CZ23 fidelity}) are in order: The seven elementary gates making up the ${\sf CZ}_{23}$ operation are optimized individually to an error of about $10^{-4}$, and then combined without any additional optimization of the composite pulse sequence or control parameters, respecting the modularity required by scalable, gate-based quantum computation. And the total intrinsic error ${\mathbb E} \equiv 1 - F = 6 \! \times \! 10^{-4}$ implied by (\ref{CZ23 fidelity}) is consistent with a linear (incoherent) accumulation of errors with number of elementary steps ${\mathbb E}_{\rm inc} \cong 7 \! \times \! 10^{-4}$, but is not consistent with a quadratic (coherent) accumulation ${\mathbb E}_{\rm coh} \cong 7^2 \! \times \! 10^{-4}.$

\subsection{Beyond QVN: {\sf CZ} gate between directly coupled superconducting qubits}
\label{directly coupled CZ section}

The main focus of this work is the {\sf CZ} gate between a directly coupled qubit and resonator in a multi-qubit QVN device. However many of our results---especially the error analysis---will also be relevant for hardware designs incorporating pairs of directly coupled superconducting qubits, the system originally considered by Strauch {\rm et al.}~\cite{StrauchPRL03}. Here we summarize the principal differences between the qubit-bus {\sf CZ} gate of Sec.~\ref{CZ gate design section} and the directly coupled qubit-qubit gate. For the latter case we assume qubits with anharmonic detunings $\eta_1$ and $\eta_2$ [defined as in (\ref{QVN hamiltonian})] and a purely transverse (off-diagonal) capacitive coupling with interaction strength $g$.

The first difference concerns a {\it frequency asymmetry} of the qubit-bus gate. Recall from Sec.~\ref{Strauch CZ gate section} that the {\sf CZ} gate is implemented by {\it decreasing} the qubit frequency $\epsilon$ from a value far above $\omega_{\rm b}$ to (approximately)
\begin{equation}
\epsilon = \omega_{\rm b} + \eta,
\label{qubit-bus crossing condition}
\end{equation}
where $\eta > 0$ is the qubit anharmonicity. By contrast, the qubit-qubit {\sf CZ} gate can be implemented either by decreasing the frequency $\epsilon_1$ of qubit 1 from a value far {\it above} that of a second qubit with a (fixed) frequency $\epsilon_2$, until 
\begin{equation}
\epsilon_1 = \epsilon_2 + \eta_1,
\label{qubit-qubit crossing condition from above}
\end{equation}
which is directly analogous to (\ref{qubit-bus crossing condition}), or from {\it below} by increasing to 
\begin{equation}
\epsilon_1 = \epsilon_2 - \eta_2.
\label{qubit-qubit crossing condition from below}
\end{equation}
The conditions (\ref{qubit-qubit crossing condition from above}) and (\ref{qubit-qubit crossing condition from below}) specify the crossings of the bare $|11\rangle$ state with $|20\rangle$ and $|02\rangle$, respectively (in the basis $|q_1 q_2\rangle$). The frequency asymmetry of the qubit-bus gate is a consequence of the harmonic spectrum of the bus and can be understood from Fig.~\ref{fig:energylevels}, which shows that when $|11\rangle$ reaches the $|02\rangle$ crossing from below, $|01\rangle$ and $|10\rangle$ also become degenerate [as expected from (\ref{qubit-qubit crossing condition from below}) when $\eta_2 \rightarrow 0$]. This would result in unwanted phase shifts of the $\overline{|01\rangle}$ and $\overline{|10\rangle}$ channels, as well as large switching errors between them.

The second major difference between the qubit-bus and qubit-qubit gates is that the additional anharmonicity in the latter case further suppresses the nonadiabatic switching errors and leads to better gate performance. This can be understood from the analysis of Sec.~\ref{fidelity estimator section}, noting that in the qubit-qubit case, adiabaticity of the $\overline{|11\rangle}$ channel is protected by an energy gap of size $\eta_1 + \eta_2 - \sqrt{2} g$, where $g$ is the qubit-qubit interaction strength. Let's estimate the $\overline{|11\rangle} \rightarrow \overline{|02\rangle}$ switching error $\mathbb{E}_2$ for a qubit-qubit {\sf CZ} gate at the upper frequency (\ref{qubit-qubit crossing condition from above}), with 
$\eta_1 = \eta_2 = 2\pi \! \times \! 300 \, {\rm MHz}$,  $g = 45 \, {\rm MHz}$, and $t_{\rm ramp}= 7 \, {\rm ns}$. In this application we use formula  (\ref{switching error formula}) with parameter values
\begin{eqnarray}
G &=& \sqrt{2} g,  \label{G value qubit-qubit case} \\ 
\Delta_{\rm on} &=& \eta_1 + \eta_2 - \sqrt{2} g, 
\label{Delta on value qubit-qubit case} \\
\Delta_{\rm off} &=& 2 \pi \times 1 \, {\rm GHz}. 
\end{eqnarray}
We note from (\ref{Delta on value qubit-qubit case}) that the anharmonicity suppressing the $\overline{|11\rangle} \rightarrow \overline{|02\rangle}$ switching error is effectively doubled in the qubit-qubit system. With these parameters we obtain $|A|^2 = 5.8 \! \times \! 10^{-6}$ and $p_{\rm sw} = 8.2 \! \times \! 10^{-8}$.

Although the $\overline{|11\rangle} \rightarrow \overline{|02\rangle}$ switching error is greatly reduced in the qubit-qubit {\sf CZ} gate, the actual gate fidelity does not fully benefit from this reduction. This is because the dominant intrinsic error in the qubit-qubit gate is the switching error $\mathbb{E}_1$ between $\overline{|01\rangle}$ and $\overline{|10\rangle}$, or the reverse, which is subdominant in the qubit-bus case (see Sec.~\ref{fidelity estimator section}).
In fact, the $\overline{|01\rangle} \leftrightarrow \overline{|10\rangle}$ switching error estimate (\ref{lower band switching error}) also applies to the qubit-qubit system (with $\eta_1 = \eta_2 = 2\pi \! \times \! 300 \, {\rm MHz}$,  $g = 45 \, {\rm MHz}$, and $t_{\rm ramp}= 7 \, {\rm ns}$), resulting in an estimated minumim fidelity of
\begin{equation}
F_{\rm min}^{({\rm est})} \equiv 1 - \mathbb{E}_1 = 1 - 2 \, p_{\rm sw} = 99.991\%.
\label{Fmin estimate for qubit-qubit CZ}
\end{equation}

\section{SYSTEM OPTIMIZATION}
\label{system optimization section}

In this section we discuss an approach for choosing optimal ${\rm QVN}_n$ device parameters. This is a complex global optimization problem that we will solve in a simple but approximate way, emphasizing the main ideas of the procedure instead of its most precise implementation.

First we consider resonator frequencies. The ${\rm QVN}_n$ processor includes $n$ memory resonators, with frequencies $\omega_{{\rm m}1} , \omega_{{\rm m}2}, \cdots $. These need to be mutually detuned (to lift degeneracies), and for simplicity we space them by $100 \, {\rm MHz}$ (a smaller value could be used). The band of memory frequencies itself needs to be well detuned from the bus to keep the idle error (to be discussed below) in check. 

Because the qubit frequency during a qubit-bus {\sf CZ} gate must approach the bus frequency from above (Sec.~\ref{directly coupled CZ section}), the bus frequency must be below the memory band. The choice of bus frequency therefore determines the lowest transition frequency that needs to be accessible by a qubit. Specifically, the qubits will need to tune $500 \, {\rm MHz}$ or so below the bus (see Fig.~\ref{mode diagram figure CZ23}). However, the minimum transition frequency may be constrained by qubit design (in addition to other considerations). In the tunable-$E_{\rm J}$ transmon, for example, this minimum frequency depends on the qubit anharmonicity $\eta$. Here we will choose a minimum qubit frequency and corresponding bus frequency appropriate for a $300 \, {\rm MHz}$ transmon. This leads to our choice of $6.5 \, {\rm GHz}$ for the bus frequency. Optimal resonator frequencies for smaller $\eta$ are unchanged, whereas for larger $\eta$ they need to be rigidly shifted upward in frequency. In particular, system frequencies for a $400 \, {\rm MHz}$ transmon will be shifted upward in frequency by about $2 \, {\rm GHz}$. Apart from this large but simple change, we expect the system optimization results, such as $g_{\rm b}$ values, to be valid for the $400 \, {\rm MHz}$ case as well. 

The frequency $\omega_{\rm off}$ can be viewed as defining a boundary between {\sf MOVE} and {\sf CZ} gates, or between consecutive {\sf MOVE} gates (see Fig.~\ref{mode diagram figure CZ23}, for example). It is also natural to perform single-qubit operations with microwave pulses at the qubit frequency $\omega_{\rm off}$. If $\omega_{\rm off}$ is too low, the error of the (approximate two-parameter) {\sf MOVE} to/from memory gate becomes significant (this is determined by the qubit-bus detuning because the dominant error is nonadiabatic leakage to the bus), whereas if $\omega_{\rm off}$ is too high the fidelity of the qubit-bus {\sf CZ} degrades (because $d\epsilon/dt$ increases). We find that $7.5 \, {\rm GHz}$ works well. At least $500 \, {\rm MHz}$ is required between $\omega_{\rm off}/2\pi$ and the memory band to keep the {\sf MOVE} errors (to/from the bus) under control. Thus we arrive at the memory frequencies given in Table \ref{parameter table}.

Having obtained prospective resonator frequencies, we turn to couplings. The most frequently used gate is expected to be the {\sf MOVE} to/from memory, which must be as fast as possible. Figure \ref{mode diagram figure CZ23} includes four examples. The gate time for this operation is approximately
\begin{equation}
\frac{\pi}{2 g_{\rm m}} + t_{\rm ramp} + 1 \, {\rm ns}.
\label{memory MOVE gate time}
\end{equation}
The first term is the $\pi$ rotation time, and the second and third are switching times (the detuning ramp can be fast because the qubit is unoccupied). Choosing $g_{\rm m}/2 \pi = 100 \, {\rm MHz}$ makes the first term $2.5 \, {\rm ns}$. It might be possible to increase $g_{\rm m}$ further, but suppresssing the resulting idling error (see below) would require an even higher empty-qubit parking frequency. The value of $t_{\rm ramp}$ is determined by the desired {\sf MOVE} gate fidelity. Because the dominant error is nonadiabatic excitation of the bus, we can estimate it using our expression 
(\ref{switching error formula}) for the switching error $p_{\rm sw}$, with 
$G = g_{\rm b}$, $\Delta_{\rm on}/2\pi  = 1.0 \, {\rm GHz}$, and $\Delta_{\rm off}/2\pi =1.8 \, {\rm GHz}$ (corresponding to $m_4$, the worst case). These values depend on our initial resonator frequency assignments. From Fig.~\ref{Asquared figure} we obtain $|A|^2 = 0.03 \, (9.8 \! \times \! 10^{-4})$ for a $1\, {\rm ns} \, (2\, {\rm ns})$ ramp. Considering the largest (worst case) value for $g_{\rm b}/2 \pi$ of $60 \, {\rm MHz}$ gives $p_{\rm sw}= 1.1 \! \times \! 10^{-4} \, (3.5 \! \times \! 10^{-6})$ for a $1\, {\rm ns} \, (2\, {\rm ns})$ ramp. Thus we conclude that the {\sf MOVE} to/from memory can be done in about $5 \, {\rm ns}$ if  $g_{\rm m}/2\pi = 100 \, {\rm MHz}$. (Here we assumed the simplest 2+1-parameter {\sf MOVE} to/from memory gate, having two pulse-shape parameters and one auxiliary $z$ rotation angle. It is also possible to implement this gate with even higher fidelity with 4+1 parameters \cite{GaliautdinovPRA12}.)

The bus coupling is found by the following ``$g$-optimization'' procedure: Consider the set of discretized $g_{\rm b}/2\pi$ values, varying from 10 to $100 \, {\rm MHz}$ in steps of $1 \, {\rm MHz}$. For each value of  $g_{\rm b}$, calculate the minimum value of $t_{\rm ramp}$ and the associated $t_{\rm gate}$ required to achieve a target fidelity, say 99.9\%. We do this by stepping through $t_{\rm ramp}$ values, estimating the fidelity using  (\ref{switching error formula}) and (\ref{finimum fidelity estimate}) from Sec.~\ref{fidelity estimator section}, which is very efficient, then confirming through a full optimization on ${\rm QVN}_1$. We then obtain, for each $g_{\rm b}$, the gate time of a 99.9\% {\sf CZ} gate, or equivalently, the function
\begin{equation}
t_{\rm gate}^{(99.9\%)}\big(g_{\rm b}\big).
\label{tgate function}
\end{equation}
The function (\ref{tgate function}) gives the time required for a {\sf CZ} gate with a given target fidelity as a function of $g_{\rm b}$. Strict $g$-optimization requires choosing $g_{\rm b}$ to minimize $t_{\rm gate}$, and this procedure leads to the best performance in any given situation. However, carrying this out leads to a different $g_{\rm b}$ for each target fidelity. Fortunately, the curvature at the minimum in (\ref{tgate function}) is small and the simpler $g_{\rm b}$ values reported in Table \ref{fidelity table}, which are independent of the target fidelity and are also constrained to be multiples of $5 \, {\rm MHz}$, lead to very little performance loss, typically 1-2$ \, {\rm ns}$ in gate time. 

Having obtained prospective resonator frequencies and couplings, we choose the empty qubit parking frequency $\omega_{\rm park}$ to control the idle error
\begin{equation}
{\mathbb E} = \big( \Omega_{\rm ZZ} t \big)^2  n^2 ,
\label{idle error definition}
\end{equation}
where $\Omega_{\rm ZZ}$ is the effective $\sigma^z \otimes \sigma^z$ coupling frequency between a memory resonator and the bus, induced by their shared qubit \cite{GaliautdinovPRA12}. The $n$-dependence in (\ref{idle error definition}) assumes the worst case. Assuming $\omega_{\rm park}/2 \pi = 10 \, {\rm GHz}$, $g_{\rm b}/2 \pi = 60 \, {\rm MHz}$ (the worst case), and $\eta/2 \pi = 400 \, {\rm MHz}$ (also the worst case)
leads to 
\begin{equation}
\frac{\Omega_{\rm ZZ}}{2\pi} = -0.881 \, {\rm kHz}.
\end{equation}
It will be necessary to keep (\ref{idle error definition}) less than the fault-tolerant threshold during a (potentially) long error correction cycle. If we assume $t=1 \, {\rm  \mu s}$, the idle error in ${\rm QVN}_4$ is $4.9 \! \times \! 10^{-4}$, which is acceptable. Reducing the parked qubit frequency to $9.5 \, {\rm GHz} \, (9.0 \, {\rm GHz})$ increases the idle error to $3.8 \! \times \! 10^{-3} \, (7.3 \! \times \! 10^{-2}).$ 

We are now able to calculate the gate time of the {\sf MOVE} to/from bus operation. The gate time is approximately
\begin{equation}
\frac{\pi}{2 g_{\rm b}} + t_{\rm ramp} + 1 \, {\rm ns},
\label{bus MOVE gate time}
\end{equation}
where the first term is between 4 and $8 \, {\rm ns}$ for the bus couplings in Table \ref{fidelity table}. As before, $t_{\rm ramp}$ is determined by the desired gate fidelity. The dominant error is nonadiabatic transition to memory, which we estimate using  (\ref{switching error formula}) with $G = g_{\rm m}$, $\Delta_{\rm on}/2\pi  = 0.5 \, {\rm GHz}$ (the worst case), and $\Delta_{\rm off}/2\pi =1.5 \, {\rm GHz}$. Note that this error is enhanced by the large value of $g_{\rm m}$. From Fig.~\ref{Asquared figure} we obtain  $|A|^2 = 5.9 \! \times \! 10^{-2} \, (1.6 \! \times \! 10^{-2} )$ for a $2\, {\rm ns} \, (3\, {\rm ns})$ ramp. Then $p_{\rm sw}= 2.4 \! \times \! 10^{-4} \, (6.4 \! \times \! 10^{-4})$ for a $2\, {\rm ns} \, (3\, {\rm ns})$ ramp. Thus we conclude that the {\sf MOVE} to/from bus takes between 7 and $12 \, {\rm ns}$, depending on the actual value of $g_{\rm b}$ and on the desired fidelity.

\begin{figure}[htb]
\centering
\includegraphics[angle=0,width=1.05\linewidth]{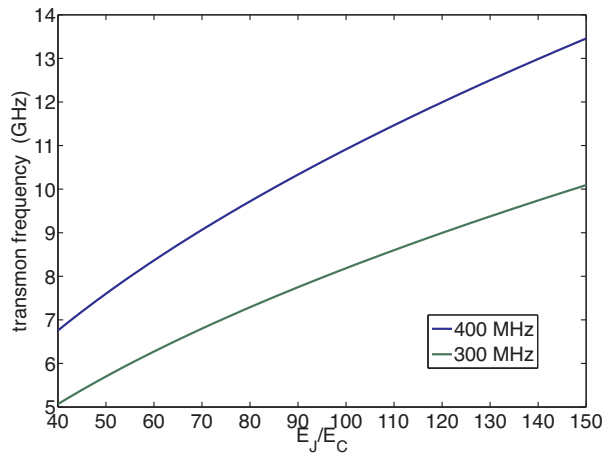}
\caption{(Color online) Transition frequency for transmon with
300 and $400 \, {\rm MHz}$ anharmonicity.}
\label{transmon frequency figure}
\end{figure}

Finally, we confirm that the assumed qubit parameters are compatible with transmons. In the large $E_{\rm J}/E_{\rm C}$ transmon regime, the qubit frequency $\epsilon$ and anharmonicity $\eta$ are given by 
\cite{KochPRA07}
\begin{equation}
\epsilon = \sqrt{8  E_{\rm J} E_{\rm C} } - E_{\rm C}
\label{transmon frequency equation}
\end{equation}
and
\begin{equation}
\eta= E_{\rm C}.
\label{transmon anharmonicity equation}
\end{equation}
We assume a split-junction flux-biased Cooper-pair box so that  $E_{\rm J}$ is tunable \cite{DiCarloNat09}. (Note that the tunable-$E_{\rm J}$ transmon is sensitive to flux noise, which will degrade $T_2$.) Combining
(\ref{transmon frequency equation}) and (\ref{transmon anharmonicity equation}) leads to the relation $\epsilon = \eta ( \sqrt{8E_{\rm J}/E_{\rm C}} - 1)$ plotted in Fig.~\ref{transmon frequency figure} for 300 and $400 \, {\rm MHz}$ anharmonicity. Because $E_{\rm J}/E_{\rm C}$ needs to be above about 50 to effectively suppress charge noise, we see that the $300 \, {\rm MHz}$ transmon can have a transition frequency as small as $5.5 \, {\rm GHz}$, whereas the $400 \, {\rm MHz}$ transmon has a minimum frequency of about $7.5 \, {\rm GHz}$. Our choice of bus frequency is indeed consistent with the $300 \, {\rm MHz}$ transmon, whereas $\omega_{\rm b}/2\pi$ and the entire spectrum of device frequencies would have to be increased by about $2 \, {\rm GHz}$ for the $400 \, {\rm MHz}$ transmon.

\section{CONCLUSIONS}\label{conclusion section}

We have investigated the problem of {\sf CZ} gate design for the superconducting QVN architecture based on a realistic, two-parameter filtered rectangular pulse (\ref{pulseEq}). The gate operates by using the non-computational qubit $|2\rangle$ state as first proposed by Strauch {\it et al.}~\cite{StrauchPRL03}. The main results are summarized in Table \ref{fidelity table}. The use of interacting eigenfunctions as computational basis states, and the use of auxiliary $z$ rotations on the qubits and bus are critical to obtaining this high performance. The ability to perform SU(2) operations on the bus resonator is nontrivial because they cannot be implemented directly with microwave pulses or frequency excursions; instead, they must be compiled into future qubit rotations.

Our investigation is limited to and constrained by the pulse profile (\ref{pulseEq}). Fidelity optimization determines the amount of time to spend switching the qubit and how long to stay at the ``on" frequency. We find that this pulse shape correctly captures the relevant pulse degrees-of-freedom for fidelities up to about 99.99\%. For higher fidelity gates, which are slower and which would require larger coupling, alternative few-parameter pulse shapes where most of the time is spent switching, could become competitive with (\ref{pulseEq}). Although we are able to obtain fidelities well above 99.99\% with profile (\ref{pulseEq}), it is possible that alternative few-parameter pulse shapes would be able to achieve the same intrinsic fidelity in less gate time.

One can also consider more complex pulse shapes with many control parameters, which can achieve nearly perfect intrinsic fidelity in a time $t_{\rm gate}$ (depending on $g_{\rm b}$) significantly shorter than obtained with pulse shape (\ref{pulseEq}). Egger {\it et al.} \cite{EggerPre12} have recently investigated this optimal control approach (using the gradient pulse shape engineering method of Khaneja {\rm et al.}~\cite{KhanejaJMR05}), and have obtained about a factor of two speedup for a qubit-qubit {\sf CZ} gate similar to that of Sec.~\ref{CZ between qubits section}. This approach clearly warrants further investigation and experimental implementation.

We did not include the effects of decoherence (or flux noise) in this work. However, an order-of-magnitude estimate of the $T_1$ decay error $\mathbb{E} \approx t_{\rm gate}/T_1$ suggests that it should be possible to demonstrate a $99.9\%$ {\sf CZ} gate with existing transmon qubits, which would be an important step towards the development of fault-tolerant quantum computation.

\begin{acknowledgments}

This work was supported by IARPA under ARO Grant No.~W911NF-10-1-0334.
It is a pleasure to thank Leonardo DiCarlo, Matteo Mariantoni, and Frank Wilhelm for useful discussions.

\end{acknowledgments}

\bibliography{/Users/mgeller/Desktop/group/publications/bibliographies/MRGpre,/Users/mgeller/Desktop/group/publications/bibliographies/MRGbooks,/Users/mgeller/Desktop/group/publications/bibliographies/MRGgroup,/Users/mgeller/Desktop/group/publications/bibliographies/MRGqc-josephson,/Users/mgeller/Desktop/group/publications/bibliographies/MRGqc-architectures,/Users/mgeller/Desktop/group/publications/bibliographies/MRGqc-general,/Users/mgeller/Desktop/group/publications/bibliographies/MRGqc-TSC,endnotes}

\end{document}